\documentclass[a4paper,twoside,openany,11pt]{memoir} 
\pdfoutput=1

\def\fullheadfoot{0}
\usepackage[english]{babel}
\usepackage{microtype,xspace}
\usepackage[utf8]{inputenc}	

\usepackage{amsfonts,amsmath,amssymb,bm,mathrsfs,mathtools,dsfont} 	
\allowdisplaybreaks 	
\usepackage{xcolor}
\usepackage{hyperref}
\usepackage{slashed}
\usepackage{multicol}

\usepackage{siunitx}
\usepackage{geometry}
\usepackage[sort&compress,numbers,merge]{natbib}
\usepackage{chapterbib}
\addto\captionsenglish{\renewcommand*{\bibname}{References}}
\makeatletter
\renewcommand{\@memb@bchap}{ 
\bibmark \prebibhook
}
\makeatother

\usepackage{graphicx}
\graphicspath{{./Figures/}}
\usepackage{caption,subcaption}
\usepackage{multirow}

\usepackage{tabularx}
\newcolumntype{Y}{>{\centering\arraybackslash}X}

\usepackage[shortlabels]{enumitem}
\setlist{itemsep=.1em,topsep=.5em}
\SetEnumerateShortLabel{i}{\textit{\roman*}}

\definecolor{red}{rgb}{0.6,.0706,.1373}
\definecolor{blue}{rgb}{0,0.396,0.741}
\colorlet{blueRef}{blue!80!black}
\colorlet{blueLink}{blue!100!black}
\hypersetup{
	colorlinks, 
	bookmarksopen, 
	bookmarksnumbered,
	citecolor=blueLink, 	
	linkcolor=blueLink,	
	urlcolor=blueLink,			
	pdftitle={Beta function ambiguities},    
	pdfsubject={RG functions in particle physics},	
	pdfauthor={Florian Herren and Anders Eller Thomsen},    	
}

\addto\captionsenglish{}

\renewcommand{\contentsname}{Contents}
\renewcommand{\printtoctitle}[1]{}

\settocdepth{subsection}
\newcommand{\app}[1][Appendices]{
	\renewcommand{\thesubsection}{\Alph{subsection}}
	\numberwithin{equation}{subsection}
	\numberwithin{figure}{subsection}
	\pagestyle{appstyle}
	\sectionlike{#1} 
}

\makeatletter
\newcommand*\ifthispageodd{%
  \checkoddpage
  \ifoddpage
    \expandafter\@firstoftwo
  \else
    \expandafter\@secondoftwo
  \fi
}
\makeatother

\usepackage[noindentafter]{titlesec} 
\counterwithout{section}{chapter} 
\counterwithout{figure}{chapter} 
\counterwithout{table}{chapter} 
\numberwithin{equation}{section} 
\setsecnumdepth{subsubsection}

\DeclareMathAlphabet{\mathsfit}{OT1}{lmss}{m}{sl}
\DeclareMathAlphabet{\mathsfbf}{OT1}{lmss}{bx}{n}
\DeclareMathAlphabet{\mathsfbfit}{OT1}{lmss}{bx}{sl}
 
\DeclareMathVersion{chaptermath}
\SetSymbolFont{operators}{chaptermath}{OT1}{cmbr}{m}{n}
\SetSymbolFont{letters}{chaptermath}{OML}{cmbrm}{b}{it}
\SetSymbolFont{symbols}{chaptermath}{OMS}{cmbrs}{m}{n}
\DeclareMathVersion{subsectionmath}
\SetSymbolFont{operators}{subsectionmath}{OT1}{cmbr}{m}{n}
\SetSymbolFont{letters}{subsectionmath}{OML}{cmbrm}{m}{it}
\SetSymbolFont{symbols}{subsectionmath}{OMS}{cmbrs}{m}{n}

\titleformat{\section}{\centering \Large \bfseries \sffamily \mathversion{chaptermath} \color{blue!90!black} }{\thesection}{15pt}{}{}
\titlespacing{\section}{0pt}{15pt}{5pt}
\titleformat{\subsection}{\large \sffamily \mathversion{subsectionmath} \color{blue!90!black} }{\thesubsection}{10pt}{}{}
\titlespacing{\subsection}{0pt}{10pt}{5pt}
\titleformat{\subsubsection}{\normalsize \sffamily \itshape \mathversion{subsectionmath} \color{blue!80!black} }{\thesubsubsection}{10pt}{}{}
\titlespacing{\subsubsection}{0pt}{10pt}{0pt}

\newcommand{\sectionlike}[1]{\phantomsection \addcontentsline{toc}{section}{#1} \sectionmark{#1}
		\begin{center}
		\needspace{8\baselineskip}
		\Large \bfseries \sffamily \mathversion{chaptermath} \color{blue!90!black} #1  
		\end{center}
	\vspace{-5pt} 
}

\ifnum\fullheadfoot=1
	\makepagestyle{standardstyle}
	\makeoddhead{standardstyle}{\sffamily \mathversion{subsectionmath} \rightmark}{}{\sffamily \bfseries{\thepage}}
	\makeevenhead{standardstyle}{\sffamily \bfseries{\thepage}}{}{\sffamily \mathversion{subsectionmath} \leftmark}
	\makeheadrule{standardstyle}{\textwidth}{\normalrulethickness}
	\makepsmarks{standardstyle}{
		\nouppercaseheads
		\createmark{section}{both}{shownumber}{\ }{\hspace{1.2em}  }
		\createmark{subsection}{right}{shownumber}{}{\hspace{1.2em} }
	}
	
	\copypagestyle{appstyle}{standardstyle}
	\makeevenhead{appstyle}{\sffamily \bfseries{\thepage}}{}{ \sffamily \mathversion{subsectionmath} \leftmark}
	\makepsmarks{appstyle}{
		\nouppercaseheads
		\createmark{section}{both}{nonumber}{\ }{\ \ }
		\createmark{subsection}{right}{shownumber}{}{\ \ }
	}
\else 
	\makepagestyle{standardstyle}
	\makeoddfoot{standardstyle}{}{\sffamily -- {\bfseries\thepage} --}{} 
	\makeevenfoot{standardstyle}{}{\sffamily -- {\bfseries\thepage} --}{}
	\copypagestyle{appstyle}{standardstyle}
\fi

\createplainmark{toc}{both}{\contentsname}
\createplainmark{bib}{both}{\bibname}

\makeatletter
\let\MyIntOrig\int
\def\MyIntSpace{\hspace{-.35em}} 
\def\int{\MyInt}
\def\MyInt{\MyIntOrig\MyIntSkipMaybe}
\def\MyIntSkipMaybe{
	\@ifnextchar_{\MyIntSkipScript}{%
		\@ifnextchar^{\MyIntSkipScript}{%
			\@ifnextchar\limits{\MyIntSkipTok}{%
				\@ifnextchar\nolimits{\MyIntSkipTok}{%
					\MyIntSpace}}}}%
}
\def\MyIntSkipScript#1#2{#1{#2}\MyIntSkipMaybe}
\def\MyIntSkipTok#1{#1\MyIntSkipMaybe}

\newcommand{\pushright}[1]{\ifmeasuring@#1\else\omit\hfill$\displaystyle#1$\fi\ignorespaces}
\makeatother

\usepackage[framemethod=TikZ]{mdframed} 
\newenvironment{theo}[1]{%
	\mdfsetup{%
			frametitle={%
				\tikz[baseline=(current bounding box.east),outer sep= 0pt, inner ysep=-.5pt]
				\node[anchor=east,rectangle,fill=blue!40] {\strut #1};}
			}%
	
	\mdfsetup{innertopmargin=2pt,linecolor=blue!40,%
		linewidth=2pt,topline=true,%
		frametitleaboveskip=\dimexpr-\ht\strutbox\relax
	}
	\begin{mdframed}[]\relax%
	}{\end{mdframed}}

\newcommand{\braces}[1]{\big\lbrace #1 \big\rbrace}
\newcommand{\brakets}[1]{\big\langle #1 \big\rangle}

\newcommand{\eminus}{\vcenter{\hbox{\scalebox{0.6}[1]{$ - $}}}}	

\newcommand{\commutator}[2]{\big[#1, \, #2 \big]}

\newcommand{\hc}{\; + \; \mathrm{H.c.} \;}
\newcommand{\andeq}{\quad \mathrm{and} \quad}

\newcommand{\dd}{\mathop{}\!\mathrm{d}}
\newcommand{\ud}[2]{\phantom{}^{#1}\phantom{}_{#2}}

\newcommand{\fdev}[1]{\dfrac{\delta}{\delta #1}}

\newcommand{\hbeta}{\beta}
\newcommand{\hgamma}{\gamma}
\newcommand{\hupsilon}{\upsilon}
\newcommand{\hrho}{\rho}
\newcommand{\sj}{\mathcal{J}}
\newcommand{\ct}{\mathrm{ct}}
\newcommand{\fscc}{\text{(FSCC)}}

\newcommand{\sscript}[1]{{\scriptscriptstyle \mathrm{#1}}}
\renewcommand{\L}{\mathscr{L}}
\newcommand{\LL}{\mathrm{L}}
\newcommand{\RR}{\mathrm{R}}
\newcommand{\U}{\mathrm{U}}
\newcommand{\SU}{\mathrm{SU}}

\newcommand{\bef}{$ \beta $-function\xspace}
\newcommand{\befs}{$ \beta $-functions\xspace}
\newcommand{\msbar}{$ \overline{\text{MS}} $\xspace}
\renewcommand{\quote}[1]{``#1''}
\newcommand{\rgfin}{RG finite\xspace}

\begin{document}

\thispagestyle{empty}
\renewcommand*{\thefootnote}{\fnsymbol{footnote}}
\vspace*{0.01\textheight}
\begin{center}
	{\sffamily \bfseries \LARGE \mathversion{chaptermath} On Ambiguities and Divergences in\\[.2em] Perturbative Renormalization Group Functions}\\[-.5em]
	\textcolor{blue!80!black}{\rule{.87\textwidth}{2.5pt}}\\
	\vspace{.02\textheight}
	{\sffamily  \mathversion{subsectionmath} \large Florian Herren$ ^{1} $\footnote{fherren@fnal.gov} 
	and Anders Eller Thomsen$ ^{2} $\footnote{thomsen@itp.unibe.ch}}\\[.3em]
	{ \small \sffamily \mathversion{subsectionmath} $ ^{1}\, $Fermi National Accelerator Laboratory, \\[-.3em]
	Batavia, IL, 60510, USA
	\linebreak $ ^{2}\, $Albert Einstein Center for Fundamental Physics, Institute for Theoretical Physics,\\[-.3em]
	University of Bern, CH-3012 Bern, Switzerland
	}
	\\[.005\textheight]{\itshape \sffamily \today}
	\\[.03\textheight]
\end{center}
\setcounter{footnote}{0}
\renewcommand*{\thefootnote}{\arabic{footnote}}%
\suppressfloats	

\begin{abstract}\vspace{-.05\textheight}
\noindent
There is an ambiguity in choosing field-strength renormalization factors in the $ \overline{\text{MS}} $ scheme starting from the 3-loop order in perturbation theory. More concerning, trivially choosing Hermitian factors has been shown to produce divergent renormalization group functions, which are commonly understood to be finite quantities. 
We demonstrate that the divergences of the RG functions are such that they vanish in the RG equation due to the Ward identity associated with the flavor symmetry. It turns out that any such divergences can be removed using the renormalization ambiguity and that the use of the flavor-improved $ \beta $-function is preferred. 
We show how our observations resolve the issue of divergences appearing in previous calculations of the 3-loop SM Yukawa $ \beta $-functions and provide the first calculation of the flavor-improved 3-loop SM $ \beta $-functions in the gaugeless limit. 

{\vspace{.7em} \footnotesize \itshape Preprint: FERMILAB-PUB-21-196-T}
\end{abstract}


\section{Introduction}
The renormalization group (RG) functions are fundamental quantities in quantum field theories (QFTs). They make it possible to relate physics at various energy scales while avoiding large logarithms in the perturbative expansion. 
The \befs are frequently used in phenomenology, but also to study the ultimate ultraviolet fates of models to ensure the absence of Landau poles~\cite{Giudice:2014tma,Litim:2014uca} or the meta-stability of the Standard Model (SM) vacuum~\cite{Bezrukov:2012sa,Degrassi:2012ry,Alekhin:2012py}. 
Additionally, the current drive for precision physics has prompted the calculation of the SM \befs to the 4-loop order for gauge and 3-loop order for Yukawa and quartic couplings~\cite{Mihaila:2012fm,Chetyrkin:2013wya,Bednyakov:2013cpa,Bednyakov:2014pia,Bednyakov:2015ooa,Zoller:2015tha,Davies:2019onf}. 

It was in connection with the 3-loop calculation of the SM matrix Yukawa \befs~\cite{Bednyakov:2014pia,Herren:2017uxn} that the ambiguity subject to our investigation made its first appearance. 
The \msbar field-strength renormalization factors are typically determined by requiring finiteness of the 2-point functions. However, starting at the 3-loop order, this prescription leaves room for an ambiguity in the form of a nontrivial unitary factor.
This is typically identified with the ambiguity inherent in taking the square root of a matrix renormalization factor.
It was recognized that the trivial choice of setting the unitary factor to the identity (taking Hermitian square roots) leads to a divergent anomalous dimension of the quark fields and the Yukawa \befs~\cite{Bednyakov:2014pia,Herren:2017uxn}. 
To be clear, when we refer to a \emph{divergent} RG function in this context, we mean that it has poles in the dimensional expansion. It is not a statement about the couplings diverging because of a Landau pole in the flow.  
At the time, the divergence problem was circumvented by choosing a suitable renormalization constant, which produced finite RG functions, while also testing that various flavor invariants were insensitive to a leftover ambiguity in the finite part of the \bef. 
With the community now at the brink of new 3- and 4-loop \bef computations, 
a more thorough understanding of the ambiguity is required to confidently perform the loop calculations and settle whether the divergences point to some deeper limitation of the regular approach. 
This discussion also directly points to an ambiguity in recent 3-loop \bef computations in Yukawa theories~\cite{Steudtner:2020tzo,Steudtner:2021fzs} and the 6-loop computation in pure scalar theory~\cite{Bednyakov:2021clp}, both of which treat theories with nontrivial flavor structure.

To elucidate the connection between the divergent RG functions and the renormalization ambiguity in more general terms, we turn to \emph{the local renormalization group} (LRG). 
This framework is, as one might expect, a local extension of the RG as most theorists are familiar with it~\cite{Shore:1986hk,Osborn:1991gm}.
The RG functions are identified as the response of local sources to a Weyl transformation of the theory. This construction is ideal for probing scale invariance and has been frequently used in the search for an $ A $-function~\cite{Osborn:1989td,Jack:1990eb,Osborn:1991gm,Jack:2013sha,Baume:2014rla,Ellwanger:2021orv}, a (monotonically) increasing function along the RG. 
The Weyl symmetry of the local theory is anomalous and the consistency of the anomaly enforces a number of nontrivial conditions on the various RG functions known as the Weyl consistency conditions~\cite{Antipin:2013sga,Jack:2013sha,Jack:2014pua,Gracey:2015fia,Poole:2019kcm}. 
A relatively recent result of the LRG is that a conformal field theory is signified not by vanishing \befs but by vanishing flavor-improved \befs, $ B = \beta - (\upsilon\, g) $, where $ \upsilon $ is an RG function related to a background gauge field of the flavor symmetry~\cite{Fortin:2012hn}.\footnote{One of the authors (AET) realized to his dismay that in the original literature $ \upsilon $ is, in fact, a Greek upsilon rather than a Latin \emph{v}~\cite{Jack:2013sha}. With this timely warning, the reader will not have to make the same mistake.} 
As such, the perturbative limit cycle found of Ref.~\cite{Fortin:2012cq} does, in fact, correspond to a zero of $ B $. 
For our present purposes, the LRG provides a comprehensive framework for working with the RG functions and their interplay with the flavor symmetry. 

In this paper we argue the following points:
\begin{enumerate}[i)]
\item the occurrence of a certain class of $ \epsilon $-poles in the RG functions is consistent with the Callan-Symanzik equation and not a sign of the theory or the renormalization scheme breaking down;
\item there is an ambiguity in choosing renormalization constants due to the flavor symmetry of a theory under a simultaneous transformation of fields and bare couplings;
\item using the ambiguity, it is always possible to remove all the poles of the class discussed in \textit{i}) simultaneously from \befs and field anomalous dimensions;
\item the flavor-improved \bef, $ B $, is unambiguous and finite and therefore a preferred choice of \bef.
\end{enumerate}
We then bridge the gap from the theoretical considerations in the LRG framework to hands-on calculations in 4-dimensional gauge-Yukawa theories, in general. In particular, we demonstrate our points \textit{i})--\textit{iv})
in the 3-loop SM computation and present the first nontrivial computation of $ B_I $ in the gaugeless limit of the SM. 

The remainder of the paper is organized as follows: In the next section, we review the LRG and introduce the majority of the notation and concepts we will need. 
In Section~\ref{sec:Renormalization_ambiguity} we discuss the renormalization ambiguity and present our main thesis. 
Section~\ref{sec:relation_to_SM} discusses what the ambiguity looks like in gauge-Yukawa theories in general and in the SM in particular. 
Finally, we conclude in Section~\ref{sec:conclusion}. 
We provide the explicit derivation of several nontrivial relations crucial to the discussion of the renormalization ambiguity and a
1-loop example showcasing the necessary computations for obtaining $\upsilon$ in the appendix.

\section{The Local Renormalization Group}
We begin with a rather long-winded review of the LRG, the purpose of which is twofold: there is a lot of notation to be introduced, not all of which is standard(ized) even among those working with the LRG; and many readers may be unfamiliar with the framework. 
We also take this opportunity to spell out the inclusion of field sources in the framework. 

\subsection{Sourcing composite operators}
Ultimately, our goal is to study the \befs and field anomalous dimensions in a dimensionally regulated 4-dimensional QFT. In the minimal subtraction (MS) scheme, these quantities are insensitive to the inclusion of relevant operators to the Lagrangian, which we will therefore ignore. Working in $ d=4- 2\epsilon $ dimensions, the generic bare action of any such QFT is 
	\begin{equation} \label{eq:flat_space_action}
	S[\Phi,\, g_0,\, \sj_0 ] = S_\mathrm{kin}[\Phi] + \! \int \dd^d x \left(g_{0,I} \mathcal{O}^I + \sj_{0,\alpha} \Phi^\alpha \right)\,,
	\end{equation}
where $ S_\mathrm{kin} $ is the kinetic term, and the scalar and fermion quantum fields are collectively denoted $ \Phi^\alpha(x)\,$. 
These matter fields are sourced by bare field sources $ \sj_{0,\alpha}(x)\,$.
While we also allow for quantum gauge fields $ A_\mu $ in the theory, we leave these implicit for the discussion in this section, as we do not wish to source them.
The last piece of the action is the interaction terms with bare couplings $ g_{0,I} $, indexed with indices $ I, J,\ldots\, $, and operators $ \mathcal{O}_I(x) $ running over all gauge-invariant marginal operators: gauge potential, Yukawa, and quartic. 

The idea behind the LRG is to allow for the computation of correlators of composite operators, as opposed to just the field correlators sourced by $ \sj_0\,$, by introducing more background sources to the theory. 
The Minkowski metric $ \eta_{\mu\nu} $ is replaced with a curved background $ \gamma_{\mu\nu}(x)\,$, which sources the energy-momentum tensor $ T_{\mu\nu} $ in the path integral. 
Accordingly, all derivatives in the action are upgraded to be covariant under diffeomorphism transformations. 
The marginal couplings, $ g_I $, are in turn upgraded to local sources $ g_I(x)\,$, so they can source the corresponding marginal operators in the path integral. 

It is also possible to source currents of global symmetries of the theory. 
We let $ G_F $ with Lie algebra $ \mathfrak{g}_F $ denote the largest global symmetry group of $ S_\mathrm{kin} $ compatible with the quantum gauge group.\footnote{By compatibility of the quantum gauge and flavor symmetry, we mean that $ [A_\mu,\,\omega]\ud{\alpha}{\beta} = 0 $ in the representation of $ \Phi^\alpha $ for any $ \omega \in \mathfrak{g}_F\,$.} 
The action of $ G_F $ on $ \Phi^\alpha $ dictates the action on $ \mathcal{O}_I $ (assumed to be covariant combinations of fields and derivatives), $ \sj\,$, and $ g_I $ so as to keep the action invariant: under the infinitesimal action of $ \omega \in \mathfrak{g}_F\,$, the sources (and fields) transforms as 
	\begin{equation} \label{eq:source_transformation}
	\begin{split}
	\delta_\omega \Phi^\alpha &= -(\omega\, \Phi)^\alpha = -\omega\ud{\alpha}{\beta} \Phi^\beta\,, \\
	\delta_\omega g_I &= - (\omega\, g)_I = g_J \omega\ud{J}{I}\,,\\
	\delta_\omega \sj_{0,\alpha} &= - (\omega\, \sj_0)_{\alpha} = \sj_{0,\beta} \omega\ud{\beta}{\alpha}\,, 
	\end{split}
	\end{equation}
employing anti-Hermitian matrix representations, $ \omega^\dagger = - \omega\,$, as is conventional in the LRG.
We will regularly use the notation $ (\omega\, \cdot) $ for the action of elements $ \omega \in \mathfrak{g}_F $ on various objects. 

The current $ J_F^{\mu} $ of the flavor symmetry is sourced by promoting $ G_F $ to a local symmetry with the background gauge field $ a_\mu(x) \in \mathfrak{g}_F $ as the source. As with any other gauge field, the infinitesimal transformation of $ a_\mu $ is given by 
	\begin{equation}
	\delta_\omega a_{\mu} = \partial_\mu \omega + \left[a_{\mu}, \omega \right] \equiv D_{\mu} \omega\,, \qquad \omega(x) \in \mathfrak{g}_F~.
	\end{equation}
It is included in the action by replacing all derivatives with $ G_F $ covariant versions such that they are now simultaneously covariant under diffeomorphism, quantum gauge symmetry, and local flavor symmetry. 

The inclusion of new sources to the theory introduces the need for new counterterms, consisting of curvature terms, covariant derivatives of the marginal couplings, as well as the field-strength tensor of the background gauge field. 
We group these into an extra action piece $ S_\ct[\gamma,\, g_0,\, a_0] $, though we do not need any details for our present purposes but rather refer to Refs.~\cite{Jack:1990eb,Jack:2013sha} for an exhaustive list of the source counterterm. 
In principle, one should also include dimension $ d-2 $ operators, $ \mathcal{O}_M^a $, as these are required for a consistent study of the Weyl anomaly even when scalar mass terms are put to zero by hand~\cite{Jack:2013sha,Baume:2014rla}. 
They will not play any direct role in our discussion, so we will ignore them along with other relevant operators, which do not play this special role. 

The action with the new sources introduced in the LRG framework is 
	\begin{equation} \label{eq:LRG_action}
	S[\Phi,\, \gamma,\, g_0,\, a_{0},\, \sj_0 ] = S_\mathrm{kin}[\Phi,\, \gamma,\, a_0] + \! \int \dd^d x \sqrt{\gamma} \left(g_{0,I} \mathcal{O}^I + \sj_{0,\alpha} \Phi^\alpha \right) + S_\ct[\gamma,\, g_0,\, a_0]\,,
	\end{equation}
with the associated vacuum functional given by
	\begin{equation}
	e^{i\mathcal{W}_0[\gamma,\, g_0,\, a_0,\, \sj_0]} \equiv \int \mathcal{D}\Phi \, \mathcal{D} A \, e^{i\,S[\Phi,\, \gamma,\, g_0,\, a_{0},\, \sj_0 ]}\,,
	\end{equation}
generating all bare connected Green's functions. Anticipating the need for renormalization, we have put subscript zeros on all sources so far, to indicate that they are \quote{bare.} 
In contrast to the conventional perturbative approach, we give up renormalizing $ \Phi $ in favor of renormalizing $ \sj_0 $. In this approach no kinetic counterterm or field rescaling is required. 
Rather, the vacuum functional is renormalized by writing each bare source as a function of finite renormalized sources, that is, 
	\begin{equation}
	\mathcal{W}[\gamma,\, g,\, a,\, \sj] = \mathcal{W}_0\big[\gamma,\, g_0(g),\, a_0(a, g),\, \sj_0(\sj, g)\big]\,,
	\end{equation}
where $ \mathcal{W} $ is the finite, renormalized vacuum functional.  

The parameterization of bare sources in terms of the renormalized couplings is ultimately a question of choosing a suitable renormalization scheme. Here we use the MS scheme (or, equivalently, \msbar upon replacing the renormalization scale with $ \mu^2= (4\pi)^{\eminus 1} \bar{\mu}^2 e^{\gamma_\mathrm{E}} $), in line with the highest order SM computations. In this scheme the counterterms exclusively remove the $ \epsilon $-poles from the Greens functions, and we can parameterize the bare sources as
	\begin{align} \label{eq:source_renormalization}
	g_{0,I} &= \mu^{k_I \epsilon} (g_{I} + \delta g_I)\,, & \sj_{0,\alpha} &= \sj_\beta Z\ud{\eminus 1\beta}{\alpha}\,, & a_{0,\mu} &= a_\mu + N^I D_\mu g_{I}\,,  \nonumber \\
	\delta g_I &= \sum_{n=1}^\infty \dfrac{\delta g_I^{(n)}}{\epsilon^n}\,, & Z &= 1 + \sum_{n=1}^\infty \dfrac{z^{(n)} }{\epsilon^n}\,, & N^I &= \sum_{n=1}^\infty \dfrac{N^{(n)I}}{\epsilon^n} \in \mathfrak{g}_F\,,
	\end{align}
where the covariant derivative working on the marginal couplings is given by 
	\begin{equation}
	D_\mu g_I = \partial_\mu g_I + (a_\mu \, g)_I\,.
	\end{equation}
$ g_I $ is made dimensionless by factoring out the renormalization scale, $ \mu\,$, to the power $ k_I\,$, which is the dimensionality of the bare couplings in $ d $ dimensions. We use the convention \emph{not} to count the index on $ k_I $ for summation conventions, i.e. only sum the index if it appears another two times. 

We proceed to discuss the various symmetries of the theory in the presence of all sources. 
Several of the important equations in the RG arise as Ward identities associated with these symmetries. 

\subsection{Weyl transformations}
In four dimensions the flat space action~\eqref{eq:flat_space_action} is scaleless at the classical level and therefore possesses a global scale symmetry parameterized by $ \sigma\,$, that is, a transformation of the metric 
	\begin{equation}
	\gamma_{\mu\nu} \longrightarrow e^{\eminus 2\sigma} \gamma_{\mu \nu}\,. 
	\end{equation} 
Invariance of the kinetic term under such a scale transformation is ensured by letting the fields transform according to their canonical dimension: 
    \begin{equation} \label{eq:Weyl_trans_fields}
    \Phi^\alpha \longrightarrow \Phi^{\sigma,\alpha} = e^{\Delta_{\alpha} \sigma } \Phi^\alpha\,. 
    \end{equation}  
With the curved-space metric of the full LRG action~\eqref{eq:LRG_action}, the scale symmetry can be localized, $ \sigma \to \sigma(x)\,$, to what is known as the Weyl symmetry.\footnote{A non-minimal coupling of scalars to the curvature is needed to ensure this local invariance.} 

In more generality, the transformation properties of the bare sources under the Weyl symmetry are dictated by their mass dimension. 
Consequently, the marginal couplings receive a nontrivial transformation when working in the $ d\neq 4 $ dimensions needed for regularization (see e.g. Ref.~\cite{Baume:2014rla}).
Symmetry of the marginal operator terms in the action is established by letting the transformation of the bare sources compensate the canonical transformation of the operators $ \mathcal{O}^I\,$, and we have
    \begin{equation} \label{eq:Weyl_trans_bare_sources}
    g_{0,I} \longrightarrow g_{0,I}^{\sigma} = e^{(d- \Delta_{\mathcal{O}^I}) \sigma} g_{0,I} = e^{k_I  \epsilon \sigma} g_{0,I}\,.
    \end{equation}
From transformation~\eqref{eq:Weyl_trans_bare_sources}, we may infer the transformation of the renormalized coupling. Using renormalization-scale independence of the bare coupling  yields
    \begin{equation}
    g_{0,I}^{\sigma} = e^{\sigma k_I \epsilon} g_{0,I}(\mu, g(\mu)) = g_{0,I}(e^{\sigma} \mu, g(\mu)) = g_{0,I}(\mu, g(\mu e^{\eminus \sigma}))\,.
    \end{equation}
The transformation law for the bare couplings, thus, implies that the renormalized couplings transform as
	\begin{equation} \label{eq:Weyl_trans_couplings}
	g_{I}(\mu) \longrightarrow g^{\sigma}_I(\mu) = g_{I}(\mu e^{\eminus \sigma}) \quad  \implies \quad \delta_\sigma g_{I} = - \sigma \dfrac{\dd g_{I}}{\dd t} = - \sigma \hbeta_{I}\,, \qquad t= \ln \mu\,, 
	\end{equation}
where $ \delta_\sigma $ is the infinitesimal response to a Weyl transformation with parameter $ \sigma\,$. This transformation rule implies that the response of the renormalized coupling to a local rescaling of space is to run to a new energy scale with the $ d $-dimensional \bef $ \hbeta\,$.

We, also, directly infer the formula for $ \hbeta_{I} $ from the bare coupling parameterization~\eqref{eq:source_renormalization}. Taking an infinitesimal Weyl transformation yields 
	\begin{equation} \label{eq:coupling_dev}
	k_I \epsilon\, \sigma \,\mu^{k_I\epsilon}(g_I + \delta g_I) = \delta_\sigma g_{0,I} = \mu^{k_I\epsilon} (\delta\ud{J}{I} + \partial^J \delta g_I) \delta_\sigma g_J\,, \qquad \partial^I \equiv \dfrac{\partial}{\partial g_I}\,.
	\end{equation}
Rearranging things a little gives
	\begin{equation} \label{eq:Weyl_bef_requirement}
	\hbeta_{I} + k_I \epsilon \, g_I +(k_I \epsilon +\hbeta_{J} \partial^J) \delta g_I = 0\,, 
	\end{equation}
which is the usual relation for the MS \bef. 

We proceed to the transformation rules for the field sources, $ \sj^\alpha\,$. The bare sources transform according to their canonical dimension, and applying the infinitesimal Weyl variation gives
	\begin{equation}
	(d- \Delta_\alpha) \sigma\, \sj_\beta Z\ud{\eminus 1 \beta}{\alpha} = \delta_\sigma \sj_{0,\alpha} = \delta_\sigma \sj_\beta Z\ud{\eminus 1\beta}{\alpha} - \delta_\sigma g_I \sj_\beta [Z^{\eminus 1} \partial^I Z \,Z^{\eminus 1}] \ud{ \beta}{\alpha}\,.   
	\end{equation}  
Plugging in the transformation of the couplings~\eqref{eq:Weyl_trans_couplings}, we recover 
	\begin{equation} \label{eq:anom_dim_def}
	\delta_\sigma \sj_{\alpha} = \sigma \, \sj_\beta \left[(d- \Delta_\alpha) \delta\ud{\beta}{\alpha} - \hgamma\ud{\beta}{\alpha} \right]\,,\qquad \hgamma\ud{\alpha}{\beta} = \hbeta_{I} [Z^{\eminus 1}  \partial^I  Z]\ud{\alpha }{\beta}\,.
	\end{equation}
$ \hgamma $ is, thus, identified with the $ d $-dimensional field anomalous dimension. 

Finally, for the background gauge field, we parameterize $ \delta_\sigma a_\mu = \partial_\mu \sigma \, \hupsilon - \sigma \hrho^I D_\mu g_I $ for $ \hupsilon(g)\,,\,\hrho^I(g) \in \mathfrak{g}_F\, $, consistent with the most general form respecting dimensionality and covariance under $ G_F\,$. 
As $ a_{0,\mu} $ is part of the covariant derivative, it must be invariant under the Weyl transformation, meaning that 
	\begin{equation}
	0 = \delta_\sigma a_{0,\mu} = \delta_\sigma a_\mu +N^I (\delta_\sigma a_\mu \, g)_I + N^I (D_\mu \delta_\sigma g)_I + \delta_\sigma g_J \partial^J N^I D_\mu g_I
	\end{equation} 
per Eq.~\eqref{eq:source_renormalization}.	
At this point, it is sufficient to plug in the transformation of the couplings and the parameterization of $ \delta_\sigma a_\mu $ to obtain 
	\begin{equation} \label{eq:upsilon_def}
	\hupsilon = N^I B_I\,, \qquad B_I = \hbeta_{I} - (\hupsilon\, g)_I
	\end{equation}
and 
	\begin{equation} \label{eq:rho_def}
	\hrho^I = -\hbeta_J \partial^J N^I - N^J \partial^I \hbeta_{J} - N^J (\hrho^I \, g)_J
	\end{equation}
from the $ \partial_\mu \sigma $ and $ \sigma $ terms, respectively. Of the two new RG functions associated with the flavor current, $ \hupsilon $ is our main interest: it bridges the gap between $ \hbeta_I $ and the flavor-improved \bef $ B_I\,$.\footnote{$ \hupsilon $ is referred to as $ S $ in parts of the literature.} 

Collecting all the infinitesimal transformation rules of the renormalized sources, the generator of an infinitesimal Weyl transformation acting on the vacuum functional is 
	\begin{multline}
	\Delta^{\! W}_\sigma = \int\dd^d x \,\bigg(2\sigma \gamma^{\mu\nu} \fdev{\gamma^{\mu\nu} } - \sigma \hbeta_{I} \fdev{g_I} + \sigma \sj_\beta \big[(d- \Delta_\alpha) \delta\ud{\beta}{\alpha} - \hgamma\ud{\beta}{\alpha} \big] \fdev{\sj_\alpha} \\
	+\left[\partial_\mu \sigma \, \hupsilon - \sigma\, D_\mu g_I \,\hrho^I \right] \cdot  \fdev{a_\mu} \bigg)\,,
	\end{multline}
where \quote{ $ \cdot $ } denotes the inner product on $ \mathfrak{g}_F\,$.
With the Weyl transformation being a classical symmetry of the action, the vacuum functional is invariant down to the Weyl anomaly: 
	\begin{equation}
	\Delta^{\! W}_{\sigma} \mathcal{W} = \int \dd^d x\, \mathcal{A}^W_\sigma(\gamma,\, g,\, a)\,.
	\end{equation} 
The anomaly, $ \mathcal{A}^W_\sigma\,$, is a local term of the sources and has been the focus of the effort to derive Weyl consistency conditions~\cite{Osborn:1991gm} and ultimately an $ A $-function in four dimensions~\cite{Jack:1990eb,Jack:2013sha,Baume:2014rla}. 

The Weyl symmetry has implications for the trace of the energy-momentum tensor. In particular, we concern ourselves with the \emph{flat-space-constant-coupling} (FSCC) limit, that is, the limit where $ \gamma_{\mu \nu} = \eta_{\mu\nu}\,$, $ g_I(x) = g_I\,$, and $ a_\mu = 0\,$, which is the phenomenologically relevant limit. 
Taking also the field sources to vanish, the Weyl symmetry implies the operator equation 
	\begin{equation} \label{eq:trace_T_1}
	[T\ud{\mu}{\mu}] = \hbeta_{I} [\mathcal{O}^{I}] + \hupsilon \cdot \partial_\mu [J_F^\mu] - \eta_a \partial^2 [\mathcal{O}^{a}_M] \qquad \fscc
	\end{equation}
for the trace of the energy-momentum tensor for Green's functions with no coinciding points.
The square brackets are taken to indicate that the operators are renormalized. 
This equation is known as the scale anomaly and signals the loss of classical scale invariance in the quantum theory. 
The trace of the energy-momentum tensor will vanish if all the RG functions---$ \hbeta_I\,$, $ \hupsilon\,$, and $ \eta_a $---do, and the theory becomes conformal. 
Note also the presence of the dimension-2 operators $ \mathcal{O}_M^{a}\,$, which is why they cannot be ignored in an analysis of the anomaly.

\subsection{Renormalization group transformations}
Osborn~\cite{Osborn:1991gm} demonstrated that the anomaly is closely related to the RG flow of $ \mathcal{W}\,$. 
Since $ \mathcal{W} $ is dimensionless, a simultaneous scaling of the all lengths and masses must leave the vacuum functional invariant, resulting in the counting identity for the physical dimension:
	\begin{equation} \label{eq:scale_generator}
	\Delta^\mu \mathcal{W} = 0\,, \qquad 
	\Delta^\mu = \mu \dfrac{\partial}{\partial \mu} + \!\int\dd^d x \left(2 \gamma^{\mu\nu} \fdev{\gamma_{\mu\nu} } + (d-\Delta_\alpha) \sj_\alpha \fdev{\sj_\alpha} \right)\,.
	\end{equation}
We can now identify the RG generator as the combination of $ \Delta^\mu $ and $ \Delta^{\! W}_\sigma $ that eliminates the derivative w.r.t.\ the metric, thus keeping lengths fixed: 
	\begin{equation}
	\Delta^{\!\sscript{RG} } = \Delta^\mu - \Delta^{\! W}_{\sigma=1} = \dfrac{\partial}{\partial t} + \!\int\dd^d x \bigg( \hbeta_{I} \fdev{g_I} + \sj_\beta \hgamma\ud{\beta}{\alpha} \fdev{\sj_\alpha} + D_\mu g_I \, \hrho^I \cdot \fdev{a_\mu} \bigg)\,.
	\end{equation}

In the FSCC limit, $ \Delta^\mu $ and $ \Delta^{\! W}_\sigma $ are both good symmetries of $ \mathcal{W}\,$. Applying $ \Delta^{\!\sscript{RG}} $ to the vacuum functional in this limit, yields the Callan-Symanzik equation 
	\begin{equation} \label{eq:CS}
	0 = \Delta^{\!\sscript{RG}} \mathcal{W} = \bigg( \dfrac{\partial }{\partial t} + \hbeta_{I} \partial^I + \!\int \dd^d x \, \sj_\beta \hgamma\ud{\beta}{\alpha} \fdev{\sj_\alpha} \bigg) \mathcal{W} \qquad \fscc\,.
	\end{equation} 
In the FSCC limit $ \Delta^{\!\sscript{RG}} \mathcal{W} = \dd \mathcal{W}\! /\!\dd t\,$, and the Callan-Symanzik equation could also have been derived following the usual arguments about invariance of the bare vacuum functional w.r.t.~$ \mu\,$. However, as we shall now see, the flavor symmetry leaves room for an ambiguity in the definition of $ \Delta^{\! W}_\sigma $ and, thus, $ \Delta^{\!\sscript{RG}} $ as well.

\subsection{Flavor transformations}
We have previously established the invariance of the action w.r.t.\ the flavor symmetry $ G_F $. In operator form, the infinitesimal generator of the symmetry is 
	\begin{equation} \label{eq:flavor_generator}
	\Delta^{\! F}_\omega = \int\dd^d x \left(D_\mu \omega \cdot \fdev{a_\mu} - (\omega \, g)_I \fdev{g_I} - (\omega \, \sj)_\alpha \fdev{\sj_\alpha} \right),\qquad \omega \in \mathfrak{g}_F\,,
	\end{equation}
which leaves $ \mathcal{W} $ invariant down to a possible flavor anomaly: 
	\begin{equation}
	\Delta^{\! F}_\omega \mathcal{W} = \int \dd^d x \, \mathcal{A}^F_\omega(\gamma,\, g,\, a)\,. 
	\end{equation}
This form of the anomaly assumes the absence of anomalies mixing with the dynamical gauge fields $ A_\mu\,$. Keren-Zur~\cite{Keren-Zur:2014sva} argued that mixed anomaly contributions with dynamical and background gauge fields are irrelevant for perturbative contributions in any event, as they are total derivatives.

The existence of the flavor symmetry introduces an ambiguity in the definition of the Weyl symmetry. In general, it holds that
	\begin{equation}
	\Delta^{\! W\,\prime}_\sigma = \Delta^{\! W}_\sigma + \Delta^{\! F}_{\sigma \alpha}\,, \qquad \alpha(g) \in \mathfrak{g}_F\,,
	\end{equation}
where $ \alpha(g) $ is a covariant function of the couplings, is an equally valid Weyl generator~\cite{Jack:2013sha,Baume:2014rla}.\footnote{Or, if the reader prefers, the generator of an equally valid Weyl transformation.} The new generator maintains 
	\begin{equation}
	\commutator{\Delta^{\! F}_{\omega}}{\Delta^{\! W\, \prime}_\sigma} = \commutator{\Delta^{\! W\, \prime}_\sigma }{\Delta^{\! W\, \prime}_{\sigma'}} =0\,,\qquad \Delta^{\! W\, \prime}_\sigma \mathcal{W} = \int\dd^d x\, \mathcal{A}^{W\, \prime}_\sigma\,,
	\end{equation} 
where the new anomaly has the same structure as the original Weyl anomaly $ \mathcal{A}^W_\sigma $, but with modified coefficients for the CP-odd terms~\cite{Keren-Zur:2014sva}. 
One way to view this ambiguity is that due to the flavor symmetry, a simple local scaling is indistinguishable from a local scaling in conjunction with a flavor rotation: physics remains unchanged.  

The flavor ambiguity in the definition of the Weyl generator corresponds to a redefinition of the RG functions, namely 
	\begin{align} \label{eq:Weyl_RG_ambiguity}
	\hbeta'_I &= \hbeta_I + (\alpha\, g)_I\,, & \hupsilon' &=  \hupsilon + \alpha\,, &
	\rho^{\prime I} &=\rho^I -\partial^I \alpha\,, &\hgamma  &= \hgamma -\alpha \,. 
	\end{align}
These functions are all equally good RG functions for any $ \alpha\,$, in that they all satisfy the Callan-Symanzik equation~\eqref{eq:CS} but with $ (\hbeta_I,\, \hgamma) \to (\hbeta'_I,\, \hgamma')\,$.\footnote{One needs simply let $ \Delta^{\!\sscript{RG}\, \prime} = \Delta^\mu - \Delta^{\! W\,\prime}_{\sigma=1} $.} 
We would, however, like to remind the reader that it is only the original set of RG function that directly corresponds to the change of couplings/renormalization constants with the renormalization scale:
	\begin{equation}
	\hbeta_{I} = \dfrac{\dd g_I}{\dd t} \andeq \hgamma = Z^{\eminus 1} \dfrac{\dd}{\dd t} Z\,. 
	\end{equation}
	
With the ambiguity in the RG functions associated to the Weyl transformation, it can be advantageous to work with the invariant objects~\cite{Fortin:2012hn}
	\begin{align} \label{eq:flavor-improved_RG_functions}
	B_I &= \hbeta_{I} - (\hupsilon\, g)_I\,, & P^I &= \hrho^I + \partial^I \hupsilon\,, & \Gamma\ud{\alpha}{\beta} = \hgamma\ud{\alpha}{\beta} + \hupsilon\ud{\alpha}{\beta}\,.
	\end{align}  
Conveniently, there is even a \quote{gauge} where these objects appear directly in the Weyl generator defined by 
	\begin{align}
	\widehat{\Delta}^W_\sigma &= \Delta^{\! W}_\sigma +\Delta^{\! F}_{-\sigma\, \hupsilon} = \Delta^{\! W\, \prime}_\sigma +\Delta^{\! F}_{-\sigma\, \hupsilon'}\\
	&= \int\dd^d x \bigg(2\sigma \gamma^{\mu\nu} \fdev{\gamma_{\mu\nu} } - \sigma B_{I} \fdev{g_I} + \sigma \sj_\beta \big[(d- \Delta_\alpha) \delta\ud{\beta}{\alpha} - \Gamma\ud{\beta}{\alpha} \big] \fdev{\sj_\alpha} 
	- \sigma\, D_\mu g_I \,P^I \cdot  \fdev{a_\mu} \bigg)\,. \nonumber 
	\end{align}
In the limit of vanishing field sources, the Ward identity associated with $ \widehat{\Delta}^{W}_\sigma $ takes the operator form 
	\begin{equation} \label{eq:trace_T_2}
	[T\ud{\mu}{\mu}] = B_{I} [\mathcal{O}^{I}] - \eta_a \partial^2 [\mathcal{O}^{a}_M] \qquad \fscc\,,
	\end{equation}
which can equivalently be obtained from Eq.~\eqref{eq:trace_T_1} by using the Ward identity associated with $ G_F $ to eliminate the current operator. 

The new trace energy-momentum identity is very important for our purposes because \emph{it guarantees the finiteness of} $ B^I\,$, which is not the case for $ \hbeta_{I}\,$. 
This observation relies on $ [T\ud{\mu}{\mu}]  $ being finite upon insertion into (non-coinciding) correlation functions. 
Meanwhile, $ \eta_a $ can be eliminated by adding the curvature term $ -\tfrac{1}{6} \eta_a R \mathcal{O}_a $ to the action, corresponding to improving the energy-momentum tensor~\cite{Baume:2014rla}. 
This leaves just the $ B_I $ term on the RHS of Eq.~\eqref{eq:trace_T_2}, and with $ [\mathcal{O}^I] $ finite, the same must hold for $ B_I\,$.

\section{Renormalization Ambiguity} \label{sec:Renormalization_ambiguity}
In this section, we discuss the source and consistency of divergences in the RG functions and how they can ultimately be removed. 

\subsection{Deriving RG functions from counterterms}
We begin our discussion of the renormalization ambiguity with a rederivation of the formulas relating the RG functions to the poles of the counterterms. 
In general, we will be concerned with $ d $-dimensional RG functions rather than their limit in $ d=4\,$, as this limit is not guaranteed to exist.
With the observation that infinities can seemingly appear in the \befs, we expand the $ d $-dimensional \bef in powers of $ \epsilon $:
	\begin{equation}
	\hbeta_I  = \dfrac{\dd g_I}{\dd t}=  \sum_{n=\eminus 1}^{\infty} \dfrac{1}{\epsilon^n} \hbeta^{(n)}_I\,,
	\end{equation}
where $ \hbeta^{(0)}_{I} $ is the standard 4-dimensional \bef.\footnote{If one were to assume that $ \hbeta $ had higher order terms in $ \epsilon\,$, one would find, as can be seen from Eq.~(\ref{eq:beta-coupling_poles}), that all higher order coefficients disappear.} 	
	
To get the \bef, one can apply a $ t $-derivative to the bare coupling parameterization~\eqref{eq:source_renormalization} to obtain Eq.~\eqref{eq:Weyl_bef_requirement}. 
Organizing the equation by powers of $ \epsilon $ then gives 
    \begin{equation} \label{eq:beta-coupling_poles}
    0 = \left[\hbeta^{(\eminus 1)}_{I} + k_I \, g_{I}\right] \epsilon + \sum_{n=0}^{\infty} \dfrac{1}{\epsilon^n} \! \bigg(\hbeta^{(n)}_{I}  + k_I \delta g_I^{(n+1)} + \sum_{k=\eminus 1}^{n-1} \hbeta^{(k)}_{J} \partial^J \delta g^{(n-k)}_I \bigg)\,.
    \end{equation}
From the $ \epsilon $ term, we conclude that $ \hbeta^{(\eminus 1)}_{I} = -k_I g_{I}\,$. Plugging this back into the equation, the term constant in $ \epsilon $ yields the key formula for the 4-dimensional \befs:
    \begin{equation} \label{eq:beta-function_formula}
	\hbeta^{(0)}_I = \left(\zeta - k_I \right) \delta g_I^{(1)}\,,
    \end{equation}
where $ \zeta = k_I g_I \partial^I $ is twice the loop-counting operator of a 2-point function (meaning that $ \zeta -k_I $ is counts twice the loop-order of $ \delta g_I $).

From Eq.~\eqref{eq:beta-coupling_poles} one also obtains recursive relations for the divergent part of the \bef, which can be determined order by order in the $ \epsilon $-poles: 
	\begin{equation}\label{eq:beta_pole_relations}
	\hbeta^{(n)}_I = (\zeta - k_I) \delta g_I^{(n+1)} - \sum_{k=0}^{n-1} \hbeta^{(k)}_{J} \partial^J \delta g_I^{(n-k)}\,, \qquad n\geq 1\,.
	\end{equation}
The traditional view holds that all the poles, $ \hbeta^{(n)} $ for $ n\geq 1 $, must vanish and, thus, that Eq.~\eqref{eq:beta_pole_relations} can be used directly as a consistency check of the calculation. 
As we have suggested, this is not always the case. 

The field anomalous dimension can be determined from the field-strength renormalization factors, $ Z $, and the \bef. In analogy with the \bef, we allow for $ \epsilon $-poles in the $ d $-dimensional anomalous dimension and expand
	\begin{equation} \label{eq:anom_dim_def_2}
	\hgamma = Z^{\eminus 1} \dfrac{\dd}{\dd t} Z= \sum_{n=0}^{\infty} \dfrac{1}{\epsilon^{n}} \hgamma^{(n)}\,.
	\end{equation} 
Seeing as $ Z $ depends on the renormalization scale only through $ g $, we can plug in the $ \epsilon $ expansion of $ Z\,$, $ \hbeta\,$, and $ \hgamma$ to obtain 
	\begin{equation} \label{eq:anomalous_dim_formula}
	0= \big(\gamma^{(0)} + \zeta z^{(1)} \big)  + \sum_{n=1}^{\infty} \dfrac{1}{\epsilon^n} \! \bigg(\hgamma^{(n)} + \zeta z^{(n+1)} + \sum_{k=0}^{n-1} \left[z^{(n-k)} \hgamma^{(k)} - \hbeta^{(k)}_{I} \partial^{I} z^{(n-k)} \right] \! \bigg)\,. 
	\end{equation}
In particular, the standard $ 4 $-dimensional anomalous dimension is given by
	\begin{equation}
	\hgamma^{(0)} = - \zeta z^{(1)}\,,
	\end{equation}
whereas the poles can be determined recursively from 
	\begin{equation} \label{eq:gamma_pole_relations}
	\hgamma^{(n)} = - \zeta z^{(n+1)} + \sum_{k=0}^{n-1} \left[ \hbeta^{(k)}_{I} \partial^{I} z^{(n-k)} - z^{(n-k)} \hgamma^{(k)} \right], \qquad n\geq 1\,.
	\end{equation}
As with the \bef, the poles are not guaranteed to vanish.

The calculation of $ \hupsilon $ from the counterterms follows lines similar to the other RG functions. Expanding the defining formula $ \eqref{eq:upsilon_def} $ in powers of $ \epsilon $ yields 
	\begin{equation} \label{eq:ups_formula_0}
	\hupsilon^{(0)} = - k_I g_I N^{(1)I}
	\end{equation}
and 
	\begin{equation} \label{eq:ups_formula_n}
	\hupsilon^{(n)} = -k_I g_I N^{(n+1)I} + \sum_{k=0}^{n-1} N^{(n-k)I} \left[ \hbeta^{(k)}_I - (\hupsilon^{(k)}\, g)_I \right], \qquad n\geq 1\,, 
	\end{equation}
which allows for recursive evaluation of the $ \hupsilon $ poles. One can derive a formula for $ \rho_I $ in a similar manner, but we will not need it for our discussions.

\subsection{Finiteness of the renormalization group equation} \label{sec:RG_finite}
Having derived relations between RG functions and counterterm poles, it is illustrative to examine how $ \epsilon $-poles can arise in the field anomalous dimension in the first place. 
Herren \emph{et al.}~\cite{Herren:2017uxn} observed that the Hermitian part of $ \hgamma $ is finite up to 3-loop orders in the SM; however, this is neither true for the anti-Hermitian part of $ \hgamma $ nor for the \befs~\cite{Bednyakov:2014pia,Herren:2017uxn}. 

The typical approach to determining the field-strength renormalization factors prescribes choosing $ Z^\dagger = Z\,$, which, obviously, yields a Hermitian $ \hgamma^{(0)}\,$. 
In this case, it follows from Eq.~\eqref{eq:gamma_pole_relations} that the anti-Hermitian part of $ \hgamma^{(1)} $ satisfies  
	\begin{equation} \label{eq:gamma_first_pole}
	\hgamma^{(1)}-\hgamma^{(1)\dagger} = \commutator{z^{(1)}}{\zeta z^{(1)}}\,.
	\end{equation}
Starting at the 3-loop order, the commutator is non-vanishing if $ \commutator{z^{(1)}_1}{z^{(1)}_2} \neq 0\,$, where $ z_\ell^{(n)} $ is used to denote the $ \ell $-loop contribution to $ z^{(n)} $. In generic theories, or indeed the SM, this commutator will not vanish.

Despite the presence of poles in the RG functions being rather unsettling, we should not dismiss the result out of hand: the RG functions are unphysical, not observables. 
Their impact on physical quantities is contained in the scaling behavior of Greens functions via the RG. Thus, it is the Callan-Symanzik equation~\eqref{eq:CS} that must be well-behaved, allowing for a smooth 4-dimensional limit. Expanding the RG functions in powers of $ \epsilon $ gives 
	\begin{equation}\label{eq:CS_eq_divs}
	\bigg( \dfrac{\partial }{\partial t} + \left(\epsilon \hbeta_{I}^{(\eminus 1)} + \beta_{I}^{(0)} \right) \! \partial^I + \int \dd^d x \, \sj_\beta \gamma\ud{(0) \beta}{\alpha} \fdev{\sj_\alpha} \bigg) \mathcal{W} 
	= - \sum_{n=1}^{\infty} \dfrac{1}{\epsilon^n} \bigg( \hbeta^{(n)}_{I} \partial^I + \int \dd^d x \, \sj_\beta \hgamma\ud{(n) \beta}{\alpha} \fdev{\sj_\alpha} \bigg) \mathcal{W}\,. 
	\end{equation}
Clearly, the LHS is finite, but any non-vanishing poles would seem to leave the RHS divergent. How, then, can we reconcile the presence of divergences in the RG functions with finite evolution of the Green's functions? 
The answer lies in the flavor symmetry. 

The flavor symmetry gives rise to a Ward identity, which can be derived by applying the generator $ \Delta^{\! F}_\omega $ of Eq.~\eqref{eq:flavor_generator} to the vacuum functional. In the FSCC limit, the anomaly vanishes and the identity reads 
	\begin{equation}
	0= -\Delta^{\! F}_\omega \mathcal{W} = \bigg(\! (\omega\, g)_I \partial^I - \! \int \dd^d x \, \sj_\beta \omega\ud{\beta}{\alpha} \fdev{\sj_\alpha} \bigg) \mathcal{W}\,. 
	\end{equation}
The flavor Ward identity and $ \hgamma^{(n)} $ being found to be anti-Hermitian in the SM lead us to posit the theorem of RG-finiteness: 
	\begin{theo}{RG-finiteness} \label{thm:RG_fin}
	The divergent part of \emph{any} set of MS/\msbar RG functions $ (\beta_I,\, \gamma) $ satisfy 
		\begin{equation*} 
		\hgamma^{(n)} \in \mathfrak{g}_F \andeq \hbeta^{(n)}_I = - \big( \hgamma^{(n)}\, g \big)_I \,, \qquad n \geq 1\,.
		\end{equation*}
	This property of the RG functions is referred to as RG-finiteness. 
	\end{theo} \vspace{-.5em}
Any \emph{RG-finite} pair $ (\beta_I,\, \gamma) $ causes the RHS of Eq.~\eqref{eq:CS_eq_divs} to vanish due to the flavor Ward identity. 
Divergences of this kind in the RG functions are, thus, consistent with finite running of the Green's functions. 
The proof that all RG functions are RG finite relies on the ambiguity of RG functions and is presented in Section.~\ref{sec:RG_function_ambiguity}. 

In theories where all couplings are invariant under $ G_F $, it is impossible to construct a non-vanishing $ \hgamma^{(n)} \in \mathfrak{g}_F\,$. In such cases, RG-finiteness implies individual finiteness of all RG functions.
This explains why higher-order calculations of RG functions in e.g. gauge-fermion theories through five loops~\cite{Luthe:2016xec,Baikov:2017ujl,Chetyrkin:2017bjc} or $ \phi^4 $ theory through six loops~\cite{Kompaniets:2017yct} has not encountered divergences of this kind.
Though it is consistent for RG-finite RG functions to feature divergences, we concede that for practical (and aesthetic) reasons, it is preferable to avoid any such. Fortunately, there is an ambiguity in choosing the counterterms that allows for choosing finite RG functions.

\subsection{Transforming the bare sources} \label{sec:Bare source transformation}
Other authors have made use of an ambiguity in the renormalization as a way to eliminate the divergent parts of the RG functions. By tweaking the field-strength renormalization constants, it has, so far, always been possible to do so. The renormalization ambiguity, thus, seems to be linked to the divergences of the RG functions.

Fortin \emph{et al.}~\cite{Fortin:2012hn} pointed out that the ambiguity in choosing $ Z $ is equivalent to performing a divergent flavor rotation; however, they only considered the impact on the finite part of the RG functions. To formalize the idea, we introduce the notion of having the flavor transformation work on the bare sources of the theory, by acting with the generator 
	\begin{equation}
	\Delta^{\! F_0}_\omega = \int\dd^d x \left(D_\mu \omega \cdot \fdev{a_{0,\mu} } - (\omega\, g_0)_I \fdev{g_{0,I}} - (\omega\, \sj_0)_\alpha \fdev{\sj_{0,\alpha}} \right),\qquad \omega \in \mathfrak{g}_F\,,
	\end{equation}
rather than the generator~\eqref{eq:flavor_generator} for rotations of the renormalized sources. 
$ \Delta^{\! F_0}_\omega $ is the generator of the classical flavor symmetry of the bare action and leaves $ \mathcal{W}_0 $ invariant down to the chiral anomaly, which is irrelevant for the perturbative computation of the field-dependent counterterms~\cite{Keren-Zur:2014sva}.\footnote{On the other hand, the chiral anomaly from the application of $ \Delta^{\! F_0}_\omega $ will effectively generate a change to the field-independent counterterms, $ S_\ct $, essential to the the perturbative $ A $-function. We will not explore this effect here.}
Whenever $ \omega $ is finite, the action of $ \Delta^{\! F_0}_\omega $ and $ \Delta^{\! F}_\omega $
coincide, as the renormalization factors and counterterms are covariant under the finite flavor transformation. 
It is when $ \omega $ is taken to have an $ \epsilon $ dependence that $ \Delta^{\! F_0}_\omega $ becomes truly interesting, as it can generate changes in the counterterms while leaving the renormalized sources unchanged. 

Here we consider flavor rotations with parameters $ u(g) \in \mathfrak{g}_F\,$ that are series of poles in~$ \epsilon\,$. 
Such rotations leave the finite part of the bare sources, i.e., the renormalized sources, unchanged while changing the counterterms, effectively generating a new set of renormalization constants.  
The same holds for finite---in the sense of non-infinitesimal---rotations given by
	\begin{equation} \label{eq:counterterm_rotation}
	U  = \exp\! \bigg[-\sum_{n=1}^{\infty} \dfrac{1}{\epsilon^n} u^{(n)}(g) \bigg]\,, \qquad u^{(n)} \in \mathfrak{g}_F\,.
	\end{equation}
The action of $ U $ on the sources follows from Eq.~\eqref{eq:source_transformation}, and the transformed sources are
	\begin{equation} \label{eq:transformed_bare_sources}
	\tilde{g}_{0,I} = (U \,g_{0})_{I}\,, \qquad \tilde{\sj}_{0,\alpha} = \sj_{0,\beta} U\ud{\dagger \beta}{\alpha}\,, \andeq \tilde{a}_{0,\mu} = U a_{0,\mu} U^\dagger + U \partial_\mu U^\dagger\,,
	\end{equation}
after the rotation, while physics remains unchanged:
	\begin{equation} \label{eq:vf_invariance}
	\mathcal{W}[\gamma,\, g,\, \sj,\, a] = \mathcal{W}_0[\gamma,\, g_0,\, \sj_0,\, a_0 ] = \mathcal{W}_0[\gamma,\, \tilde{g}_0,\, \tilde{\sj}_0,\, \tilde{a}_0 ]\,,
	\end{equation} 
down to the chiral anomaly. 
We will take $ U $ to be a flavor-covariant function of $ g_I $ so as not to spoil the flavor symmetry under finite $ G_F $ transformations. 
Since $ \sj_\alpha\,$, the finite part of the bare source, remains unchanged, the effect of the transformation is to multiply the field-strength renormalization by a unitary factor:
	\begin{equation}
	\tilde{Z}\ud{\alpha}{\beta} = U\ud{\alpha}{\gamma} Z\ud{\gamma}{\beta}\,.
	\end{equation} 
This is why the ambiguity is typically identified with a unitary ambiguity related to taking the square root of $ Z^\dagger Z $~\cite{Jack:1990eb,Bednyakov:2014pia,Herren:2017uxn}. 

It is by no means an accident that this ambiguity enters at the 3-loop order in the SM. 
The $ u^{(n)}(g) $ are anti-Hermitian 2-point tensor structures---flavor-covariant polynomials in the couplings. 1PI tensor structures of this type are only possible beginning at the 3-loop order in completely general gauge-Yukawa theories~\cite{Poole:2019kcm}. 
Furthermore, the 1-loop 2-point tensor-structures all commute, ensuring that the leading contribution to $ u^{(n)}(g) $ is 3-loop order. Hence, all complications from the ambiguity can be ignored in 2-loop computations.
The possibility of choosing a nontrivial contribution to $ u^{(n)} $ already at the 2-loop order in the SM, as pointed out in Refs.~\cite{Bednyakov:2014pia,Herren:2017uxn}, seems to ignore flavor-covariance and was extraneous to removing any divergences in $ \gamma\,$.

The renormalization ambiguity is directly relevant to the parameterization of general gauge-Yukawa theories.
If one does not fix the ambiguity with the Hermitian choice $ Z= Z^\dagger $, it is generally possible to add anti-Hermitian 1PR tensor structures to $ z^{(1)} $ with the $ U $ rotation.   
This translates directly into 1PR tensors in the finite part of $ \gamma $ and $ \beta_I $. 
Similarly, examining the structure of $ N^I $ required to cancel the divergences of the Green's functions outlined in Sec.~\ref{sec:SM_NI_calculations}, one will find that also the simple pole of $ N^I $ can only contain 1PR tensor structures if $ z^{(1)} $ is not Hermitian. 
We conclude that the parameterization of the finite parts of the RG functions presented in Ref.~\cite{Poole:2019kcm} are valid only when the ambiguity is fixed such that $ z^{(1)} $ is Hermitian.  

There is some room for differing interpretations of the renormalization ambiguity. Since the $ U $ rotation leaves the renormalized vacuum functional and sources unchanged, it is our point of view that it parameterizes a class of equally valid counterterms. It is not really a renormalization scheme change, since the renormalized couplings remain unchanged under the transformation. The only consequence of the transformation that we are aware of is a change in the RG functions.

\subsection{Ambiguity in the RG functions} \label{sec:RG_function_ambiguity}
As we have seen, the RG functions are determined by the poles of the counterterms. Thus, the renormalization ambiguity, in the form of the flavor rotation $ U\,$, induces an ambiguity in the RG functions. 
The full derivation of the transformation of the RG functions under a $ U $ rotation can be found in App.~\ref{app:RG_transformation}, whereas here we just refer the result:
	\begin{align}
	\Delta \gamma &\equiv \tilde{\gamma} - \hgamma =  - \hbeta_I \,U \partial^I  U^\dagger\,, \label{eq:Del_gamma}\\
	\Delta \beta_I &\equiv \tilde{\beta}_I - \hbeta_I = - (\Delta \gamma\, g)_I\,, \label{eq:Del_beta} \\
	\Delta \upsilon & \equiv \tilde{\upsilon} - \hupsilon = - \Delta \gamma\,, \label{eq:Del_upsilon}
	\end{align} 
where $ \Delta \gamma \in \mathfrak{g}_F\,$. 
For completeness, we have also determined that 
	\begin{equation}
	\Delta \rho^I \equiv \tilde{\rho}^I - \hrho^I = \partial^I \Delta \gamma\,.
	\end{equation} 

At finite order in $ \epsilon\,$, Eq.~\eqref{eq:Del_gamma} gives
	\begin{equation}
	\Delta \gamma^{(0)} = \zeta u^{(1)}\,.
	\end{equation}
with the parameterization~\eqref{eq:counterterm_rotation} for the rotation. 
Similarly, all the poles, $ \Delta \gamma^{(n)} $, contain a term $ \zeta u^{(n+1)} $ from the $ \hbeta^{(\eminus 1)} $ contribution, while the remaining terms only depend on the poles $ u^{(k)}, k \leq n\,$. 
$ \zeta $ is the loop-counting operator, so by carefully choosing the terms $ u^{(n+1)} $, one can engineer any covariant $ \Delta \gamma \in \mathfrak{g}_F\,$.
Hence, as long as the original RG functions were \rgfin, it is always possible to transform the counterterms such as to make $ \tilde{\gamma} $ finite, as was found to be the case in the SM~\cite{Bednyakov:2014pia,Herren:2017uxn}.

From the transformation of the \bef~\eqref{eq:Del_beta}, it further follows that \emph{\hyperref[thm:RG_fin]{RG-finite} RG functions will remain RG finite under the renormalization transformation}. 
Conversely, RG functions that were not \rgfin cannot be made so by applying the transformation. 
In particular, when $ \tilde{\gamma} $ is made finite starting from RG-finite RG functions, then $ \tilde{\beta}_I $ will be too. 

Turning to the $ \hupsilon $ transformation rule~\eqref{eq:Del_upsilon}, we observe that the change in $ \hupsilon $ is opposite to the change in $ \hgamma\,$. Accordingly, the unambiguous, or flavor-improved, RG functions $ B_I $ and $ \Gamma $ of Eq.~\eqref{eq:flavor-improved_RG_functions} are invariant not only under a redefinition of the Weyl transformation but also under the transformation of the renormalization constants. 
In fact, from comparing the ambiguity in the definition of the Weyl transformation~\eqref{eq:Weyl_RG_ambiguity} to the change of the RG functions under the counterterm transformation, it follows that it is possible to change the counterterms in such a manner as to make the RG functions agree with any one of the possible Weyl transformations. 
The two ambiguities are closely related although not quite the same: one is an ambiguity in defining the Weyl symmetry, while the other is an ambiguity in choosing counterterms.   

The last observation we wish to make regarding the RG transformation is that it is always possible to choose the counterterms such that $ \tilde{\upsilon} =0\,$. 
This is always the case as $ \hupsilon \in \mathfrak{g}_F\,$, and we can, therefore, find a $ U $ such that $ \Delta \gamma = \hupsilon\,$. With this choice,  the flavor-improved RG functions coincide with the regular ones; $ \tilde{\beta}_{I} = B_I $ and $ \tilde{\gamma} = \Gamma\,$.

We are now a position to prove \hyperref[thm:RG_fin]{RG-finiteness} of all RG functions. Using the renormalization ambiguity, we choose RG functions $ (B_I,\, \gamma) $ (for instance, it is always possible to choose $ (\beta_I,\, \gamma) = (B_I,\, \Gamma) $), ensuring finiteness of the \bef according to the discussion around Eq.~\eqref{eq:trace_T_2}. If $ \gamma $ can be made finite up to arbitrarily high loop order without changing the associated $ \beta_I = B_I\,$, then the RG functions can be made separately finite to any loop order, establishing RG-finiteness of the RG functions. 

We assume for contradiction that there is a smallest loop order $ \ell > 0$ such that no $ (B_I,\, \gamma) $ is finite up to order $ \ell\,$. Then there exists a $ \gamma $ such that its leading contribution to the divergent  part $ \gamma_\ell $ is $ \ell $-loop order.
Finiteness of the Callan--Symanzik equation~\eqref{eq:CS_eq_divs} reduces to   
\begin{equation} \label{eq:Gamma_CS}
	K_\epsilon \! \int \dd^d x \, \sj_\beta \gamma_\sscript{div} \ud{\beta}{\alpha} \fdev{\sj_\alpha} \mathcal{W}= 0\,,
\end{equation}
where $ \gamma_\sscript{div} = K_\epsilon \gamma\,$. 
$ K_\epsilon $ is an operator that extracts the divergent piece of its operand.
The leading contribution to the 2-point functions in Eq.~\eqref{eq:Gamma_CS} is given between the free-field propagators and $ \gamma_\ell\,$, providing the constraint
\begin{equation}
	\gamma_\ell = -  \gamma_\ell^{\dagger} \implies \gamma_\ell \in \mathfrak{g}_F\,.
\end{equation}
The leading contribution to several of the $ n $-point functions in Eq.~\eqref{eq:Gamma_CS} are given by the couplings $ g_I $ dressed with free propagators and $ \gamma_\ell\,$. 
Thus,\footnote{While we have no proof, it seems to us that it might well be impossible to construct a flavor-covariant $ \gamma_\ell(g) \in \mathfrak{g}_F $ satisfying $ (\gamma_\ell\, g)_I = 0 $ in fully general gauge-Yukawa theories. In that eventuality, $ \gamma = \Gamma $ must be finite along with $ B_I\,$.} 
\begin{equation} \label{eq:gamma_ell_action}
	(\gamma_\ell\, g)_I =0\,.
\end{equation}
It follows that the renormalization ambiguity can be utilized to construct a new set of RG functions $ (B_I,\, \tilde{\gamma}) $, choosing $ \Delta \gamma = \gamma_\ell\,$, which leaves $ \tilde{\beta}_I = B_I $ by Eqs.~\eqref{eq:Del_beta} and~\eqref{eq:gamma_ell_action}. 
Clearly, $ \tilde{\gamma} $ is finite up to and including the $ \ell $-loop order, for contradiction. Accordingly, all RG functions are RG finite.

\section{Gauge-Yukawa Theories} \label{sec:relation_to_SM}
After the somewhat abstract discussions in the previous sections, we exemplify the discussions with the SM as a concrete example of a gauge-Yukawa theory and show by explicit calculation that the RG functions are indeed \rgfin, as we demonstrated that they must be. 

\subsection{Generic gauge-Yukawa theories} \label{sec:gY_formalism}
We consider the most general class of renormalizable QFTs in four dimensions: gauge-Yukawa theories.
We leave out all relevant couplings, as they do not impact any of the RG functions discussed in the present paper in the \msbar scheme.\footnote{Based on the arguments in the previous section, we expect the RG functions of the relevant couplings to suffer from similar ambiguities.}
The Lagrangian of completely generic gauge-Yukawa theories can be put into the form 
	\begin{multline}\label{eq:gy_Lagrangian}
	\L =  -\tfrac{1}{4} a^{\eminus 1}_{0,AB} F^{A}_{\mu\nu} F^{B\mu\nu} + \tfrac{1}{2} (D_\mu \phi)_{a} (D^{\mu} \phi)_a + i \psi_i^{\dagger} \bar{\sigma}^\mu (D_\mu \psi)^i + J_a Z_{ab}^{\eminus 1} \phi_b + \big( \eta_i Z^{\eminus 1} \ud{i}{j} \psi^j \hc \big) \\
	- \tfrac{1}{2} \phi_a \big(y_{0,aij} \psi^i \psi^j \hc \big)  - \tfrac{1}{24} \lambda_{0,abcd} \phi_a \phi_b \phi_c \phi_d + \L_\mathrm{ghost} + \L_\text{gauge-fixing} 
	\end{multline}
by collecting all scalars in a scalar multiplet $ \phi_a $ and writing the fermions as a Weyl spinor multiplet $ \psi^i $. 
The gauge fields of the gauge group $ G $ (a product group with any number of simple or Abelian factors) are packed into a single multiplet $ A^{A}_\mu $ with collected index $ A $ belong to the adjoint representation of $ G\,$. 
The bare gauge couplings $ a_{0,AB} $ are put on the gauge kinetic term to emphasize the gauge couplings having the structure of a two-index matrix. 
For the full details of the construction, we refer the reader to Ref.~\cite{Poole:2019kcm}.    
$ y_{0,aij} $ and $ \lambda_{0,abcd} $ are the bare Yukawa and quartic couplings, respectively, while $ \sj_\alpha = (\eta_i,\, J_a) $ is the collection of field sources with field-strength renormalization factors $ Z_{ab} $ and $ Z\ud{i}{j}\,$.

To connect the gauge-Yukawa theory to the notation employed elsewhere in the paper, the marginal couplings and operators are identified with
	\begin{equation}
	g_{0,I} = \braces{a_{0,AB},\, y_{0,aij},\, \lambda_{0,abcd}} \andeq \mathcal{O}^{I} = \braces{-\tfrac{1}{4} a_{0,BC}^{\eminus 2} F^{A}_{\mu\nu} F^{C\mu\nu}, \, -\tfrac{1}{2} \phi_a \psi^i \psi^j,\, -\tfrac{1}{24} \phi_a \phi_b \phi_c \phi_d}\,.
	\label{eq::gIOI}
	\end{equation}
In $ d $-dimensions, the mass dimensions of the bare couplings are given by $ k_I = \braces{2,\, 1,\, 2}\,$. 
For the matter fields and their normalization factors, they are
	\begin{equation}
	\Phi^\alpha = \braces{\psi^i,\, \phi_a} \andeq Z\ud{\alpha}{\beta} = \braces{Z\ud{i}{j},\, Z_{ab}}\,,
	\end{equation}
where $ Z\ud{\alpha}{\beta} $ is understood to be a block diagonal matrix. 

The flavor symmetry $ G_F $ is the largest symmetry group of the kinetic terms compatible with the gauge symmetry $ G\,$.
In a theory with $ N_f $ Weyl fermions and $ N_s $ real scalars, it satisfies $ G \times G_F \subseteq \U(N_f) \times \mathrm{O}(N_s) $, ensuring that $ \commutator{A_\mu}{a_\nu} =0\,$.\footnote{Assuming, of course, that all product gauge groups have some matter charged under them. Were this not the case, the corresponding gauge fields would completely decouple from the rest of the theory.}  
The transformation properties of the fields under a group element $ h\in G_F\,$, $ \phi_a \to h_{ab} \phi_b $ and $ \psi^i \to h\ud{i}{j} \psi^j\,$, determine the transformation of the couplings and renormalization factors. 
In particular, the Yukawa and quartic couplings transform as 
	\begin{equation}
	y_{aij} \to h_{ab} y_{bk\ell} (h^\dagger)\ud{k}{i} (h^\dagger)\ud{\ell}{j} \andeq \lambda_{abcd} \to h_{ae} h_{bf} h_{cg} h_{dh} \lambda_{efgh}\,.  
	\end{equation}
Due to the scalar fields being in a real representation, the elements satisfy $ h_{ab} = h^\dagger_{ba}\,$. The quantum gauge fields, on the other hand, do not transform under the flavor symmetry, leaving the coupling $ a_{AB} $ invariant. 

The action of an element $ \omega \in \mathfrak{g}_F $ can be ascertained by considering the infinitesimal transformation $ h = 1 - \omega\,$, in line with the anti-Hermitian representation of the Lie algebra employed in Eq.~\eqref{eq:source_transformation}. 
The action is defined by the field transformation such that
	\begin{equation}
	(\omega\, \phi)_a = \omega_{ab} \phi_b \andeq (\omega\, \psi)^i = \omega\ud{i}{j} \psi^j\,.
	\end{equation}
The coupling indices, meanwhile, transform in the conjugate representations (reality of the scalar field representation ensures that $ \omega_{ab} = -\omega_{ba} $):
	\begin{equation}
	\begin{split}
	(\omega \, y)_{aij} &= \omega_{ab} y_{bij} - y_{akj} \omega\ud{k}{i} - y_{aik} \omega\ud{k}{j}\,,\\
	(\omega\, \lambda)_{abcd} &= \omega_{ae} \lambda_{ebcd} + \omega_{be} \lambda_{aecd} + \omega_{ce} \lambda_{abed} + \omega_{de} \lambda_{abce}\,.
	\end{split}
	\end{equation}  
Finally, the \hyperref[thm:RG_fin]{RG-finiteness} condition reads 
	\begin{equation}
	\begin{split}
	\hbeta^{(n)}_{aij} &= - \hgamma^{(n)}_{ab} y_{bij} + y_{akj} \hgamma\ud{(n)k}{i} + y_{aik} \hgamma\ud{(n)k}{j}\,, \\
	\hbeta^{(n)}_{abcd} &= -\hgamma^{(n)}_{ae} \lambda_{ebcd} - \hgamma^{(n)}_{be} \lambda_{aecd} - \hgamma^{(n)}_{ce} \lambda_{abed} - \hgamma^{(n)}_{de} \lambda_{abce}
	\end{split}
	\end{equation} 
in gauge-Yukawa theories. 

\subsection{Flavor symmetry of the SM}
The SM admits an $ \SU(3)_q \times \SU(3)_u \times \SU(3)_d \times \SU(3)_\ell \times \SU(3)_e \times \U(1)^5 $ flavor symmetry of the kinetic term,\footnote{One of the six $ \U(1) $ phases of the fields is identified with $ \U(1)_Y $} under which only the Yukawa couplings, 
	\begin{equation}
	\L \supset - y_u\ud{i}{j} \overline{Q}_{\LL,i} \tilde{H} U^j_{\RR} -  y_d\ud{i}{j} \overline{Q}_{\LL,i} H D^j_{\RR} - y_e\ud{i}{j} \overline{L}_{\LL,i} H E^j_{\RR} \hc\,, 
	\end{equation}	
transform non-trivially. Here, $\overline{Q}_{\LL}$ and $\overline{L}_{\LL}$ are the left-handed quark and lepton doublet, whereas $U_{\RR}$, $D_{\RR}$ and  $E_{\RR}$ are the right-handed up-type quark, down-type quark and
lepton singlets.
$H$ and $\tilde{H}$ denote the Higgs doublet and its charge conjugate.

Keeping with the non-Hermitian representation for $ \mathfrak{g}_F\,$, the infinitesimal transformation of the fields and Yukawas under said symmetry is given by 
	\begin{align}
	\delta_\omega Q^{i} &= - \omega_q\ud{i}{j} Q^{j}\,, &\delta_\omega U^{i} &= - \omega_u\ud{i}{j} U^{j}\,, & \delta_\omega D^{i} &= - \omega_d\ud{i}{j} D^{j}\,, \nonumber \\
	\delta_\omega L^{i} &= - \omega_\ell\ud{i}{j} L^{j}\,, & \delta_\omega E^{i} &= - \omega_e\ud{i}{j} E^{j}\,, & \delta_\omega H & = -\omega_h H
	\end{align}
($ \omega_h  $ is a purely imaginary number) and
	\begin{align} 
	\delta_\omega y_u\ud{i}{j} &= -(\omega \, y_u)\ud{i}{j} =-\omega_q\ud{i}{k} y_u\ud{k}{j} + y_u\ud{i}{k} \omega_u\ud{k}{j} - \omega_h y_u\ud{i}{j}\,, \label{eq::deltaomegayu}\\
	\delta_\omega y_d\ud{i}{j} &= -(\omega \, y_d)\ud{i}{j} = -\omega_q\ud{i}{k} y_d\ud{k}{j} + y_d\ud{i}{k} \omega_d\ud{k}{j} + \omega_h y_d\ud{i}{j}\,, \label{eq::deltaomegayd} \\
	\delta_\omega y_e\ud{i}{j} &= -(\omega \, y_e)\ud{i}{j} = -\omega_\ell\ud{i}{k} y_e\ud{k}{j} + y_e\ud{i}{k} \omega_e\ud{k}{j} + \omega_h y_e\ud{i}{j}\,,
	\end{align}
for $ \omega \in \mathfrak{g}_F\,$. 
The subscript on $ \omega $ or any other elements of $ \mathfrak{g}_F\,$, e.g., the generators $ t^{P} $, are used to denote the representation (effectively fundamental representations of the $ \SU(3) $ product groups). 
All other couplings and fields are in the trivial representation of $ G_F\,$, i.e., they do not transform. 

Introducing the background gauge field $ a_\mu = a^{P}_\mu t^{P} \in \mathfrak{g}_F\,$, the kinetic term of the SM is 
	\begin{multline}
	\L_\mathrm{kin} = i \overline{Q}_\LL \gamma^\mu (D_\mu + a_\mu) Q_\LL + i \overline{U}_\RR \gamma^\mu (D_\mu + a_\mu) U_\RR + i \overline{D}_\RR \gamma^\mu (D_\mu + a_\mu) D_\RR \\
	+ i \overline{L}_\LL \gamma^\mu (D_\mu + a_\mu) L_\LL + i \overline{E}_\RR \gamma^\mu (D_\mu + a_\mu) E_\RR +|(D_\mu + a_\mu)H|^2\,, 
	\end{multline}
where $ D_\mu $ is the ordinary gauge-covariant derivative.
The flavor current is determined by taking an $ a_\mu^{P} $ derivative, giving 
	\begin{equation} 
	J_F^{P\mu} = i \overline{Q}_\LL \gamma^\mu t^{P}_q Q_\LL  + i \overline{U}_\LL \gamma^\mu t^{P}_u U_\RR + i \overline{D}_\RR \gamma^\mu t^{P}_d D_\RR + i \overline{L}_\LL \gamma^\mu t^{P}_\ell L_\LL + i \overline{E}_\RR \gamma^\mu t^{P}_e E_\RR - t_h^{P} H^\dagger \overleftrightarrow{D}^\mu H\,.
	\label{eq::JF}
	\end{equation} 
The flavor indices have been kept implicit here to avoid unnecessary clutter. Once again, $ t^{P}_h $ are just charges, that is, imaginary numbers.

\subsection{Divergences in the SM RG functions}
As mentioned several times already, the poles of the SM anomalous dimensions have previously been subject to some consideration~\cite{Bednyakov:2014pia,Herren:2017uxn}, though how the poles of the anomalous dimension reflect in the \befs has not previously been considered in detail. 
Using the 3-loop counterterms $ Z^\dagger Z $ and $ \delta g_I $ provided by Ref.~\cite{Herren:2017uxn}, we derive the divergent parts of the RG functions and verify that they are indeed \rgfin. 

Since renormalization only fixes the combination $ Z^{\dagger} Z\,$, we fix the renormalization ambiguity with the Hermitian choice $ Z = Z^{\dagger}\,$. As pointed out in Section~\ref{sec:RG_finite}, this choice immediately removes any anti-Hermitian contribution to $ \hgamma^{(0)} $ for all the fields:\footnote{The fields are in one-to-one correspondence with the representation of $ G_F $. We use the same indices for the field anomalous dimensions as for the representations despite the finite part, $ \gamma_f^{(0)} $, not being the representation of an element of $ \mathfrak{g}_F $; however, it does make the notation more intuitive for the divergent.}
	\begin{equation}
	\hgamma^{(0)}_f - \hgamma^{(0)\dagger}_f = 0\,, \qquad f\in \braces{q,u,d,\ell,e}\,.
	\end{equation}
With the choice of Hermitian $ Z $, we find that $ \hgamma^{(n)}_{h} = 0 $ for $ n \geq 1 $ using the counterterms of Ref.~\cite{Herren:2017uxn}. Since $ H $ does not carry a flavor index, $ \commutator{z^{(j)}_h}{\hgamma^{(k)}_h} = 0 $ for all $ j,k $ and the anti-Hermitian part of $ \hgamma_h^{(n)} $ vanishes by ~\eqref{eq:gamma_first_pole} along with generalizations to higher poles.
For the RG functions to be RG finite $ \hgamma_h^{(n)} $ must therefore vanish in agreement with explicit computation. 

Starting at the 3-loop order, we find nontrivial contributions to the quark anomalous dimensions given by 
	\begin{align}
	(4\pi)^6 \hgamma^{(1)}_q &= \dfrac{a_1}{192} \commutator{y_u y_u^\dagger}{y_d y_d^\dagger} + \dfrac{1}{64} \commutator{y_u y_u^\dagger y_u y_u^\dagger}{y_d y_d^\dagger} + \dfrac{1}{64} \commutator{y_d y_d^\dagger y_d y_d^\dagger}{y_u y_u^\dagger}\,, \label{eq::polesstart}\\
	(4\pi)^6 \hgamma^{(1)}_u &= \dfrac{1}{32} y_u^\dagger \commutator{y_d y_d^\dagger}{y_u y_u^\dagger} y_u\,, \\
	(4\pi)^6 \hgamma^{(1)}_d &= \dfrac{1}{32} y_d^\dagger \commutator{y_u y_u^\dagger}{y_d y_d^\dagger} y_d\,,
	\end{align}
where $ a_1 = g_1^2 $ is the (squared) hypercharge gauge coupling.
As expected, the poles are anti-Hermitian in agreement with the hypothesis $ \gamma^{(1)} \in \mathfrak{g}_F\,$. The first poles of the quark Yukawa \befs are also extracted from the counterterms giving 
	\begin{align}
	(4\pi)^6 \hbeta_{y_u}^{(1)} =\,&  - \dfrac{a_1}{192} \commutator{y_u y_u^\dagger}{y_d y_d^\dagger} y_u - \dfrac{1}{64} \commutator{y_u y_u^\dagger y_u y_u^\dagger}{y_d y_d^\dagger} y_u \nonumber\\
	&- \dfrac{1}{64} \commutator{y_d y_d^\dagger y_d y_d^\dagger}{y_u y_u^\dagger} y_u + \dfrac{1}{32} y_u y_u^\dagger \commutator{y_d y_d^\dagger}{y_u y_u^\dagger} y_u\,, \\
	(4\pi)^6 \hbeta_{y_d}^{(1)} =\,&  - \dfrac{a_1}{192} \commutator{y_u y_u^\dagger}{y_d y_d^\dagger} y_d - \dfrac{1}{64} \commutator{y_u y_u^\dagger y_u y_u^\dagger}{y_d y_d^\dagger} y_d \nonumber\\
	&- \dfrac{1}{64} \commutator{y_d y_d^\dagger y_d y_d^\dagger}{y_u y_u^\dagger} y_d + \dfrac{1}{32} y_d y_d^\dagger \commutator{y_u y_u^\dagger}{y_d y_d^\dagger} y_d\,.
	\end{align}
The similarities between $ \hgamma^{(1)}_f $ and $ \hbeta_y^{(1)} $ are quite evident when organized like this, and it is a simple matter to verify that $ \hbeta^{(1)}_y = - (\hgamma^{(1)} \, y) $ in agreement with \hyperref[thm:RG_fin]{RG-finiteness}.
The $ \epsilon^2 $-poles of the RG functions are found to be 
	\begin{align}
	(4\pi)^6 \hgamma^{(2)}_q &= - \dfrac{a_1}{32} \commutator{y_u y_u^\dagger}{y_d y_d^\dagger} + \dfrac{3}{32} \commutator{y_u y_u^\dagger y_u y_u^\dagger}{y_d y_d^\dagger} + \dfrac{3}{32} \commutator{y_d y_d^\dagger y_d y_d^\dagger}{y_u y_u^\dagger}\,, \\
	(4\pi)^6 \hgamma^{(2)}_u &= - \dfrac{3}{16} y_u^\dagger \commutator{y_d y_d^\dagger}{y_u y_u^\dagger} y_u\,, \\
	(4\pi)^6 \hgamma^{(2)}_d &= - \dfrac{3}{16} y_d^\dagger \commutator{y_u y_u^\dagger}{y_d y_d^\dagger} y_d\,,
	\end{align}
and 
	\begin{align}
	(4\pi)^6 \hbeta_{y_u}^{(2)} =\,&  +\dfrac{a_1}{32} \commutator{y_u y_u^\dagger}{y_d y_d^\dagger} y_u - \dfrac{3}{32} \commutator{y_u y_u^\dagger y_u y_u^\dagger}{y_d y_d^\dagger} y_u \nonumber\\
	&- \dfrac{3}{32} \commutator{y_d y_d^\dagger y_d y_d^\dagger}{y_u y_u^\dagger} y_u - \dfrac{3}{16} y_u y_u^\dagger \commutator{y_d y_d^\dagger}{y_u y_u^\dagger} y_u\,, \\
	(4\pi)^6 \hbeta_{y_d}^{(2)} =\,&  +\dfrac{a_1}{32} \commutator{y_u y_u^\dagger}{y_d y_d^\dagger} y_d - \dfrac{3}{32} \commutator{y_u y_u^\dagger y_u y_u^\dagger}{y_d y_d^\dagger} y_d \nonumber\\
	&- \dfrac{3}{32} \commutator{y_d y_d^\dagger y_d y_d^\dagger}{y_u y_u^\dagger} y_d - \dfrac{3}{16} y_d y_d^\dagger \commutator{y_u y_u^\dagger}{y_d y_d^\dagger} y_d\,. \label{eq::polesend}
	\end{align}
They also satisfy RG-finiteness, while higher-order poles cannot appear at the 3-loop order. Meanwhile, for the lepton sector, we find that all RG poles vanish identically:  
	\begin{equation}
	\hgamma_{\ell,e}^{(n)} = 0\,, \qquad  \hbeta_{y_e}^{(n)} = 0\,, \qquad n\geq 1\,.   
	\end{equation}
This is due to $ y_e $ being the only coupling to carry $ \SU(3)_{\ell,e} $ quantum numbers, making it impossible to construct a flavor-covariant anti-Hermitian combination of the couplings. 
Of course, it is well-known that the SM allows for choosing $ y_e $ diagonal with the right flavor rotations. 
If, on the other hand, one were to include right-handed neutrinos with a Higgs coupling analogous to that of the up-quarks, anti-Hermitian combinations become possible, and the RG poles of the lepton sector would presumably mirror those of the quark sector.   
All in all, explicit computations demonstrate that fixing $ Z=Z^\dagger $ in the SM leads to RG-finite RG functions up to the 3-loop order.

\subsection{SM calculation of the $ N^I $ counterterm}\label{sec:SM_NI_calculations}
The flavor RG functions are determined by the $ N^I $ counterterm of the background gauge field, cf. Eq.~\eqref{eq:source_renormalization}. 
Although $ N^I $ counterterms are needed at the 1-loop order, $ \upsilon $ is only nontrivial from the 3-loop order~\cite{Poole:2019kcm}.
For a direct computation of $N^I$ in the SM, we need to compute the 3-loop counterterms for renormalized 1PI Green's functions of the flavor current $J_F^{P\mu}$ for all fermionic fields in the
SM.\footnote{These corrections contain subdivergences associated with corrections to the coupling of the current with the scalar doublet, hence we also have to compute these corrections up to the 2-loop order.} Said
Green's functions are related to $N^I$ by
	\begin{equation} \label{eq:flavor_Greens_function}
	\big\langle [\Phi^\alpha](x) [J_F^{P\mu}](y) [\Phi_\beta^\dagger](z) \big\rangle_{\! \sscript{1PI}}= \big(\delta^{PQ} + N^{QI} (t^P\, g)_I \big)  Z\ud{\dagger \alpha}{\gamma} \big\langle \Phi^\gamma(x) J_F^{Q\mu}(y) \Phi_\delta^\dagger(z) \big\rangle_{\! \sscript{1PI}} Z\ud{\delta}{\beta}\,,
	\end{equation}
using square brackets to denote renormalized operators and going to the FSCC limit. Using different fermion fields for $ \Phi^\alpha\,$, we can isolate the various contributions to the flavor current~\eqref{eq::JF} and corresponding representations of $ N^I\,$. 

For the computation of the 3-loop Green's functions, we have to extend the computational setup employed in Ref.~\cite{Davies:2019onf}. To this end, we introduce the field $a_\mu$ as a non-dynamical degree of freedom
and treat $t_f^P$ as non-commuting matrices similar to the Yukawa couplings. All relevant Feynman diagrams are generated using \texttt{QGRAF}~\cite{Nogueira:1991ex}
and are subsequently transformed into \texttt{FORM}~\cite{Ruijl:2017dtg} expressions as well as mapped onto integral families, using \texttt{Q2E} and \texttt{EXP}~\cite{Harlander:1997zb,Seidensticker:1999bb}.
We use the \texttt{COLOR}~\cite{vanRitbergen:1998pn} package for the computation of color factors and evaluate the 1- to 3-loop integrals using \texttt{FORCER}~\cite{Ruijl:2017cxj} after nullifying one of the external momenta.

Having obtained the 3-loop Green's functions expressed with bare parameters, we employ the coupling and field-strength counterterms computed in Ref.~\cite{Herren:2017uxn} to express them in terms of renormalized quantities. 
The remaining divergences are absorbed with the current counterterms allowing for the determination of $ N^I $.
One remark concerning the renormalization of $t_f^P$  is in order: for the actual computation, it is advantageous
to treat them in the same way as Yukawa matrices, i.e., to extract renormalization constants for each of them from the various contributions to $\big\langle [\Phi^\alpha] [J_F^{P\mu}] [\Phi_\beta^\dagger] \big\rangle_{\! \sscript{1PI}}$ in the same manner as the Yukawa matrices in Ref.~\cite{Herren:2017uxn}.
We can then match the counterterms of $t_f^P$ to an Ansatz for all possible Yukawa matrix structures in $N_f^{I} (t^P\, g)_I $ and are, thus, able to extract the counterterm $N^I\,$.
An explicit example of the procedure at the 1-loop order is presented in Appendix~\ref{app::NI}.

While the flavor-current approach captures all contributions to $ N^I $ present in pure Yukawa theories, it is insufficient for gauge theories.
The issue is that the gauge couplings are singlets under the flavor symmetry, meaning that $ (t^{P} \, a)_{AB} = 0\,$. Clearly, any such contribution drops from the Green's functions~\eqref{eq:flavor_Greens_function}, and the corresponding parts of $ N^I $ cannot be determined with that approach. Up to 3-loop order in there is only one such term, in $ N^I $ given by\footnote{It is convenient to parameterize $ N^{I} $ with a dummy variable $ \hat{g}_I $ rather than using indices and Kronecker deltas. } 
	\begin{equation} \label{eq:a1_NI_contribution}
	N_q^{I} \hat{g}_I \supset \dfrac{n_{a_1}}{(4\pi)^6} \hat{a}_1 \commutator{y_u y_u^\dagger }{y_d y_d^\dagger} 
	\end{equation} 
for some $ n_{a_1} $. 
The combined requirement that $ N^I $ is anti-Hermitian and that there is a flavor-covariant underlying parameterization for the general theory~\eqref{eq:gy_Lagrangian} leaves just four possible terms at the 3-loop order similar to the four gauge contribution in the general $ \upsilon $ parameterization~\cite{Poole:2019kcm}.
In the SM this leaves only the one term~\eqref{eq:a1_NI_contribution}.\footnote{This is not to say that there are no other gauge contributions in $ N^I \hat{g}_I $ but that these come with $ \hat{y}^{(\dagger)}_f $ and can be obtained with the flavor-current approach.}   
 
An alternative way of determining $ N^I $, which should address \emph{all} contributions is by considering Green's function of the renormalized marginal operators $[\mathcal{O}^I]$ from Eq.~(\ref{eq::gIOI}):
	\begin{equation}
	\begin{split}
	\big\langle [\Phi^\alpha]&(x) [\mathcal{O}^I](y) [\Phi_\beta^\dagger](z) \big\rangle_{\! \sscript{1PI}} \\
	= \,& 	\partial^I g_{0,J} \,Z\ud{\dagger \alpha}{\gamma} \brakets{\Phi^\gamma(x) \mathcal{O}^J(y) \Phi_\delta^\dagger(z) }_{\! \sscript{1PI}} Z\ud{\delta}{\beta} 
	- Z\ud{\dagger \alpha}{\gamma}\brakets{\Phi^\gamma(x) N^I \cdot \partial_\mu J_F^\mu(y) \Phi_\delta^\dagger(z) }_{\! \sscript{1PI}} Z\ud{\delta}{\beta}\\
	& + \delta^d(y-x) \partial^I Z\ud{\dagger \alpha}{\gamma}\brakets{\Phi^\gamma(x) \Phi_\delta^\dagger(z) }_{\! \sscript{1PI}} Z\ud{\delta}{\beta} 
	+\delta^d(y-z) Z\ud{\dagger \alpha}{\gamma} \brakets{\Phi^\gamma(x) \Phi_\delta^\dagger(z) }_{\! \sscript{1PI}} \partial^I Z\ud{\delta}{\beta}\,,
	\end{split}
	\end{equation}
where, again, $\partial^I$ denotes coupling derivatives as per Eq.~\eqref{eq:coupling_dev}. 
Taking this coupling to be the hypercharge coupling of the SM requires us to compute Green's functions with insertions of
the $\U(1)$ field-strength tensor, as well as momentum-dependent insertions of derivatives of bare coupling constants w.r.t.\ the $\U(1)$ gauge coupling. 
While a similar computation was performed in Ref.~\cite{Fortin:2012hn} for the case of a Yukawa theory,\footnote{Yukawa theories could be done with the flavor-current approach in a straightforward manner.} it requires careful routing of momenta through individual diagrams and cannot be achieved by simply nullifying one of the external momenta.
Consequently, we refrain from explicitly calculating this contribution.

\subsection{Flavor-improved RG functions in the SM}
We are now in a position to present $\hupsilon$ in the gaugeless limit of the SM after applying formulas~\eqref{eq:ups_formula_0} and ~\eqref{eq:ups_formula_n} to the $ N^I $ counterterms. 
Following Eqs.~\eqref{eq::deltaomegayu} and~\eqref{eq::deltaomegayd} the action of $ \hupsilon $ on the Yukawa couplings is given by 
	\begin{align}
	(\hupsilon \, y_u)\ud{i}{j} = \hupsilon_{q}\ud{i}{k} y_{u}\ud{k}{j} - y_{u}\ud{i}{k} \hupsilon_{u}\ud{k}{j} \andeq
	(\hupsilon \, y_d)\ud{i}{j} = \hupsilon_{q}\ud{i}{k} y_{d}\ud{k}{j} - y_{d}\ud{i}{k} \hupsilon_{d}\ud{k}{j} 
	\end{align}
in terms of the representation on the three quark flavor product groups.\footnote{We find that $ \hupsilon_\ell= \hupsilon_e =0 $, as one would expect.} 
We find $\hupsilon_f$ are given by
\begin{align}
(4\pi)^6 \hupsilon_q &= \left(\frac{3}{32\epsilon^2} + \frac{1}{64\epsilon} - 1\right) \left( \commutator{y_u y_u^\dagger}{y_dy_d^\dagger y_dy_d^\dagger} + \commutator{y_dy_d^\dagger}{ y_uy_u^\dagger y_uy_u^\dagger } \right),\\
(4\pi)^6 \hupsilon_u &= \left(\frac{3}{16\epsilon^2} - \frac{1}{32\epsilon} + \frac{1}{8}\right) y_u^\dagger \commutator{y_d y_d^\dagger}{y_u y_u^\dagger} y_u~,\\
(4\pi)^6 \hupsilon_d &= \left(\frac{3}{16\epsilon^2} - \frac{1}{32\epsilon} + \frac{1}{8}\right)y_d^\dagger \commutator{y_uy_u^\dagger}{y_dy_d^\dagger} y_d~.
\end{align}
Our results for $\hupsilon$ exactly cancel the divergent contributions to the anomalous dimensions and \befs in Eqs.~(\ref{eq::polesstart}--\ref{eq::polesend}), hence explicitly rendering the
flavor-improved RG functions, $ B_I $ and $ \Gamma $, finite. 
Finiteness of the flavor-improved RG functions can also be used to fix the poles of the $ a_1 $ contribution to $ \upsilon $ once gauge couplings are reintroduced. 
This does, however, not constrain the more interesting finite part.
 
The parameterization of $ \hupsilon $ in the generic gauge-Yukawa theory~\eqref{eq:gy_Lagrangian} has 6 free parameters for the fermion representation and 3 for the scalar and is provided in Ref.~\cite{Poole:2019kcm}. 
The finite part of $\hupsilon$ in the SM uniquely fixes two of the pure-Yukawa fermion coefficients:
	\begin{equation}
	\mathfrak{f}_4^{(3)} = -\tfrac{3}{8} \andeq \mathfrak{f}_5^{(3)} = -\tfrac{5}{16}\,.  
	\end{equation}  
The Weyl consistency conditions derived from Osborn's Equation~\cite{Poole:2019kcm} with input from a (currently) partial reconstruction of the generic 4--3--2 \befs~\cite{Davies:2021}, provides the independent constraint  
	\begin{align}
	\mathfrak{f}_4^{(3)} - 4\mathfrak{f}_5^{(3)}= \tfrac{7}{8}\,,
	\end{align}
which is satisfied by our result and serves to validate our computation. Interestingly, with this input and full knowledge of the 4--3--2 \bef coefficients, the 3-loop $ \hupsilon $ will be fully determined by the Weyl consistency conditions.

\section{Conclusion} \label{sec:conclusion}
In this paper, we exhaustively discuss the origin of, and resolution to, divergences and ambiguities in perturbative RG functions. 
Our discussion reveals that the divergences of RG functions in the dimensional regulator, first observed in 3-loop SM computations full Yukawa matrix dependence, are generic to all models with nontrivial flavor structure.
We show that said divergences are related to $G_F$, the global flavor symmetry of the kinetic terms of the matter fields. Our investigations show that the flavor symmetry protects the theory, and, in spite of the appearance of explicit divergences in individual RG functions, the RG flow is finite.
Finiteness of the Callan-Symanzik equation is guaranteed by the $ G_F $ Ward identity and \hyperref[thm:RG_fin]{RG-finiteness} implies that the divergences of field anomalous dimensions are elements of the Lie algebra of $ G_F $ \emph{and} their action on the couplings produce the divergences of the associated \befs.
We have shown this to be the case in \emph{all} four-dimensional QFTs and expect similar cancellations to hold in any spacetime dimension.

We then examined the divergences in the anomalous dimensions of the quark fields and Yukawa \befs in the SM. 
These RG functions were obtained using the prescription of Hermitian square roots of the field renormalization constants.
We found that, indeed, the divergences are consistent with RG-finiteness. 
Through a diagrammatic 3-loop computation we obtained the counterterm of the background gauge field of $ G_F\,$, which was promoted to a local symmetry.
This counterterm allowed us to compute the shift from divergent \befs to the flavor improved \bef $B$ in the SM.
The computation of $B$ in the SM, along with the Weyl consistency conditions, will allow us to generalize this shift to \emph{all} gauge-Yukawa theories.

In general, all RG functions depend on the prescription used for the square roots of field-strength renormalization constants. To apply the results presented here one
needs to choose all square roots of renormalization constants to be Hermitian unless dealing with the flavor-improved quantities.

While the ambiguity in RG functions does not affect physical observables, as it is related to field redefinitions, there are several scenarios where it might affect phenomenological applications of RG functions:
\begin{enumerate}[i)]
\item RG evolution of couplings using divergent \bef (on account of it being ill-defined);
\item matching of Yukawa or scalar quartic couplings in effective field theories if not performed via physical observables;
\item study of the RG evolution of Yukawa couplings with textures that are not protected by additional symmetries;
\item numerical RG evolution with a \bef choice leading to numerical instabilities.
\end{enumerate}
As a consequence, we recommend the use of the flavor-improved \bef, $ B_I\,$, which is guaranteed to be finite. 

\subsection*{Acknowledgments}
AET would like to thank Hugh Osborn and Colin Poole for helpful discussions and comments on the work. 
The work of AET has received funding from the Swiss National Science Foundation (SNF) through the Eccellenza Professorial Fellowship ``Flavor Physics at the High Energy Frontier'' project number 186866.
This document was prepared using the resources of the Fermi National Accelerator Laboratory (Fermilab), a U.S. Department of Energy, Office of Science, HEP User Facility. Fermilab is managed by Fermi Research Alliance, LLC (FRA), acting under Contract No. DE-AC02-07CH11359.

\app

\subsection{Transformation of RG functions under the renormalization ambiguity} \label{app:RG_transformation}
This appendix contains the derivation of the response of the RG functions under the counterterm transformation discussed in Sec.~\ref{sec:Bare source transformation}.

\subsubsection{Alternative counterterm parameterization}
Before deriving the variation of the RG functions under the divergent flavor rotation, it is useful to consider a different parameterization of the coupling counterterms in place of the more direct parameterization employed in Eq.~\eqref{eq:source_renormalization}. 
We employ the compact notation $ Z\ud{I}{J} $ to denote the tensor product of field-strength renormalization factors with open indices to match the couplings. For gauge-Yukawa theories, we have explicitly
\begin{equation} \label{eq:Z_coupling_indices}
	Z\ud{I}{J} = \braces{\delta_{AC} \delta_{BD},\, Z_a\ud{ij}{bk\ell}, \, Z_{abcd,efgh} } = \braces{\delta_{AC} \delta_{BD}, \,Z_{ab} Z\ud{i}{k} Z\ud{j}{\ell}, \,Z_{ae} Z_{bf} Z_{cg} Z_{dh} }\,.
\end{equation}
This form assumes the use of the background-field gauge, which maintains a gauge symmetry preventing the renormalization of the gauge field itself~\cite{DeWitt:1967ub,Abbott:1980hw}. 
In other gauges, many additional counterterms are needed for the various gauge interactions and a $ Z_{AB} \neq \delta_{AB} $ would appear in $ Z\ud{I}{J}\,$. 

We extract a factor of the field-strength renormalization from the bare couplings such that  
	\begin{equation} \label{eq:counterterm_parameterzation}
	\mu^{\eminus k_I \epsilon} g_{0,I} = (g_J + \delta h_J) Z\ud{\eminus 1 J}{I}\,,
	\end{equation}
This parameterization is similar to what one would use, having renormalized the fields $ \Phi^\alpha $ rather than the field sources. Thus, $ \delta h_I $ is identified as the counterterm of the 1PI vertex contribution itself. 
Interestingly, under the counterterm transformation~\eqref{eq:transformed_bare_sources}, where $ Z\ud{I}{J} \to \tilde{Z}\ud{I}{J} = U\ud{I}{K} Z\ud{K}{J} $ and $ g_{0,I} \to \tilde{g}_{0,I} = g_{0,J} U\ud{\dagger J}{I}\,$, the $ \delta h_I $ counterterm is invariant, as follows from 
	\begin{equation}
	\mu^{k_I \epsilon} (g_I + \delta h_I) =  g_{0,J} Z\ud{J}{I} \longrightarrow  \tilde{g}_{0,J} \tilde{Z}\ud{J}{I} = g_{0,J} Z\ud{J}{I}\,.
	\end{equation}   

The \bef formula in terms of the new parameterization can be derived by applying a $ t $ derivative to Eq.~\eqref{eq:counterterm_parameterzation} as per usual. With the $ t $-independent bare couplings this yields\footnote{For the coupling dimension it holds that $ k_I Z\ud{I}{J}= k_J Z\ud{I}{J}\,$.}
	\begin{equation}
	0 = \left[k_J \epsilon (g_J + \delta h_J) + \hbeta_J + \hbeta_K \partial^K  \delta h_J - (g_K + \delta h_K) \hgamma\ud{K}{J} \right] Z\ud{\eminus 1 J}{ I}\,.
	\end{equation}
Consequently, 
	\begin{equation} \label{eq:beta_formula_2}
	\hbeta_{I} = -\epsilon k_I (g_I + \delta h_I) - \hbeta_{J} \partial^J \delta h_I + (g_J + \delta h_J) \hgamma\ud{J}{I}\,,
	\end{equation}	
from which $ \hbeta $ can be determined order by order in $ \epsilon\,$. This form is particularly useful for determining the effect of the divergent flavor rotation on $ \hbeta\,$, since $ g_I $ and $ \delta h_I $ are invariants.

\subsubsection{Transformation of the ordinary RG functions}
Given the transformation of the renormalization constants induced by the divergent rotation $ U $~\eqref{eq:counterterm_rotation}, we wish to determine the associated changes in the RG functions, $ \Delta \beta_I = \tilde{\beta}_I - \hbeta_{I} $ and $ \Delta \gamma = \tilde{\gamma} - \hgamma\,$. 
We will begin with the \emph{assumption} that $ \Delta \gamma \in \mathfrak{g}_F\,$, and end up showing that this is consistent with the transformation. 
All quantities are flavor-covariant functions of the renormalized couplings on account of the $ G_F $ symmetry. Any flavor-covariant function $ f(g)\,$ satisfies
	\begin{equation} \label{eq:flavor_covariance}
	(\omega\, f) = (\omega\, g)_I \partial^I f\,, \qquad \omega \in \mathfrak{g}_F\,,
	\end{equation}
that is, the transformation of $ f $ is what one gets from transforming all the couplings that make up $ f\,$. This holds regardless of the what open indices $ f $ might have.
 
Since the vertex counterterm is invariant, $ \delta\tilde{h}_I = \delta h_I\,$, the change of the \bef is most easily determined from Eq.~\eqref{eq:beta_formula_2}, giving 
	\begin{equation}
	\Delta \beta_I = - \Delta \beta_{J} \partial^J \delta h_I + (g_J + \delta h_J) \Delta \gamma\ud{J}{I}\,.
	\end{equation}
Under the assumption $ \Delta \gamma \in \mathfrak{g}_F\,$, it holds that $ \delta h_J \Delta \gamma\ud{J}{I} = - (\Delta \gamma \, \delta h)_I\,$.  Eq.~\eqref{eq:flavor_covariance} is, thus, recast as
	\begin{equation}
	\Delta \beta_I = -(\Delta \gamma\, g)_I - \left[\Delta \beta_{J} + (\Delta \gamma \, g)_J \right] \partial^J \delta h_I\,, 
	\end{equation}
which is clearly solved by 
	\begin{equation} \label{eq:beta_change}
	\Delta \beta_I = -(\Delta \gamma \, g)_I\,.
	\end{equation}
Hence, the change of the \bef is determined entirely by the change of the field anomalous dimension. 

Next, we determine the change in the field anomalous dimension in terms of $ U $. Using formulas~\eqref{eq:anom_dim_def_2} for the definition of the anomalous dimension and~\eqref{eq:beta_change} for the transformation of the \bef, the transformation of the anomalous dimension is 
	\begin{equation}
	\begin{split}
	\Delta \gamma &= Z^{\eminus 1} U^\dagger \tilde{\beta}_I \partial^I U Z + Z^{\eminus 1} \Delta \beta_I \partial^I Z \\
	& = U^\dagger \tilde{\beta}_I \partial^I U + Z^{\eminus 1} \commutator{U^\dagger \tilde{\beta}_I \partial^I U }{Z} - Z^{\eminus 1} (\Delta \gamma\, g)_I \partial^I Z
	\end{split}
	\end{equation}
Under a regular flavor transformation, $ Z $ transforms like the adjoint representation, i.e. $ (\omega \, Z) = \commutator{\omega}{Z}\,$. Furthermore, $ U^\dagger \partial^I U $ is a Maurer--Cartan form and, so, takes values in $ \mathfrak{g}_F $. Hence, 
	\begin{equation}
	\Delta \gamma= U^\dagger \tilde{\beta}_I \partial^I U + Z^{\eminus 1} \big( [U^\dagger \tilde{\beta}_J \partial^J U -\Delta \gamma]\, g\big)_I \partial^I Z\,,
	\end{equation}
which is solved by
	\begin{equation}\label{eq:gamma_change_prelim}
	\Delta \gamma = U^\dagger \tilde{\beta}_I \partial^I U\,.
	\end{equation}
Crucially, we observe that $ \Delta \gamma \in \mathfrak{g}_F $ in line with our original assumption. 
While already sufficient to determine $ \Delta \gamma $ iteratively in the $ \epsilon $-poles, further simplifications of Eq.~\eqref{eq:gamma_change_prelim} can be obtained using flavor covariance~\eqref{eq:flavor_covariance} to write $ (\Delta \gamma\, g)_I \partial^I U = \commutator{\Delta \gamma}{U}\,$.
The transformation properties of $ U $ is due to it being the representation of an element of $ G_F\,$, thereby inheriting the transformation $ (\omega\, U) = \commutator{\omega}{U}\,$. 
With a rearrangement of the equation, one then finds 
	\begin{equation} \label{eq:gamma_change}
	\Delta \gamma = - \hbeta_I \,U \partial^I  U^\dagger\,. 
	\end{equation}
This establishes a direct relation between $ \Delta\gamma $ and the divergent rotation $ U\,$.

\subsubsection{Transformation of the flavor RG functions}
We will also want to determine the change $ \Delta \upsilon = \tilde{\upsilon} - \hupsilon $ and $ \Delta \rho = \tilde{\rho} - \hrho\,$ of the flavor RG functions under the divergent flavor rotation $ U\,$.
Both $ \hupsilon $ and $ \hrho $ are determined by the $ N^I $ counterterm of the flavor background field $ a_{0,\mu}\,$. 
From Eq.~\eqref{eq:transformed_bare_sources}, it follows that
	\begin{equation}
	\tilde{a}_{0,\mu} = a_\mu + U\commutator{a_\mu}{U^{\dagger}} + U\partial^I U^{\dagger} \partial_\mu g_I + U N^{I} U^{\dagger} D_\mu g_I\,.
	\end{equation}
While $ a_\mu $ is finite and, so, invariant under the transformation, it follows from covariance of $ U $ that the change of $ N^I $ is given by  
	\begin{equation}
	\Delta N^I = \tilde{N}^{I} - N^I = U \commutator{N^I}{U^\dagger} + U \partial^I U^\dagger\,.
	\end{equation}
The transformation of $ \hupsilon $ then follows from the definition \eqref{eq:upsilon_def}:
	\begin{equation}
	\begin{split}
	\Delta \upsilon &= \Delta B_I \tilde{N}^I + B_I \Delta N^I = \Delta B_I \tilde{N}^I + U \commutator{\hupsilon }{U^\dagger} + \hbeta_I U \partial^I U^\dagger -(\hupsilon\, g)_I U \partial^I U^\dagger\\
	&= -\big((\Delta \gamma + \Delta \upsilon)\, g \big)_I \tilde{N}^I - \Delta \gamma\,,
	\end{split}
	\end{equation}
utilizing flavor covariance of various quantities and Eq.~\eqref{eq:gamma_change}.	
Evidently, this equation is solved with
	\begin{equation}
	\Delta \upsilon = - \Delta \gamma\,,
	\end{equation}
which determines the transformation properties of $ \hupsilon\,$.

Finally, to determine the change of $ \hrho\,$, we begin from the definition~\eqref{eq:rho_def}:
	\begin{equation}\label{eq:Del_rho_1}
	\begin{split}
	\Delta \rho^I =& - \hbeta_J \partial^J \Delta N^I - \Delta N^J \partial^I \hbeta_J - \Delta N^J (\rho^I\, g)_J \\
	& -\Delta \beta_J \partial^J \tilde{N}^I - \tilde{N}^J \partial^I \Delta \beta_J - \tilde{N}^J (\Delta \rho^I \, g)_J\,.
	\end{split}
	\end{equation}
Rather a lot of algebra is needed at this stage to bring things to a reasonable manageable form.  
The second line can be evaluated using, covariance of 
	\begin{equation}
	(\omega\, N^{I}) = \commutator{\omega}{N^I} + \omega\ud{I}{J} N^J \,,
	\end{equation}
where $ \omega\ud{I}{J} $ is the representation of the Lie algebra produced from $ (\omega \, g)_I = - g_J \omega\ud{J}{I}\, $. 
The second line of Eq.~\eqref{eq:Del_rho_1} evaluates to 
	\begin{equation} \label{eq:Del_rho_line_2}
	(\Delta \gamma \, \tilde{N}^I)+ \tilde{N}^{J} \partial^I (\Delta \gamma\, g)_J - \tilde{N}^J (\Delta\rho^I \, g)_J = 
	\commutator{\Delta \gamma}{\tilde{N}^I} + \tilde{N}^J \big((\partial^I \Delta \gamma - \Delta \rho^I)\, g \big)_J\,.
	\end{equation}
The first line, meanwhile, evaluates to 
	\begin{equation} \label{eq:Del_rho_line_1}
	\begin{split}
	- \hbeta_J \partial^J &\Delta N^I - \Delta N^J \partial^I \hbeta_J - \Delta N^J (\rho^I\, g)_J \\
	=& - \left( \hbeta_J \partial^J (U \partial^I U^\dagger) + \partial^I \hbeta_J (U \partial^J U^\dagger) \right) - \hbeta_J \left( \partial^J U \commutator{N^I}{U^\dagger} + U \commutator{N^I}{\partial^J U^\dagger} \right) \\
	&- U \commutator{\hbeta_J \partial^J N^I + N^J \partial^I \hbeta_J + N^J (\hrho^I \, g)_J + \hrho^I}{U^\dagger}\,,
	\end{split}
	\end{equation}
the entire last term of which vanishes by the definition of $ \hrho^I\,$. 
The two first terms of Eq.~\eqref{eq:Del_rho_line_1} evaluate as  
	\begin{equation}
	\begin{split}
	- \left( \hbeta_J \partial^J (U \partial^I U^\dagger) + \partial^I \hbeta_J (U \partial^J U^\dagger) \right)
	&= \partial^I \Delta \gamma + \hbeta_J \big( \partial^I U \partial^J U^\dagger - \partial^J U \partial^I U^\dagger \big) \\
	&= \partial^I \Delta \gamma - \commutator{\Delta \gamma}{U \partial^I U^\dagger}
	\end{split}
	\end{equation}
and 
	\begin{equation}
	\begin{split}
	- \hbeta_J \left( \partial^J U \commutator{N^I}{U^\dagger} + U \commutator{N^I}{\partial^J U^\dagger} \right) &= - \Delta \gamma \, U \commutator{N^I}{U^\dagger} + U \commutator{N^I}{U^\dagger \Delta \gamma}\\
	&= - \commutator{\Delta \gamma}{N^I} - \commutator{\Delta \gamma}{U\commutator{N^I}{U^\dagger}}\,,
	\end{split}
	\end{equation}
respectively. Altogether then, the first line of Eq.~\eqref{eq:Del_rho_1} reduces to
	\begin{equation}
	- \hbeta_J \partial^J \Delta N^I - \Delta N^J \partial^I \hbeta_J - \Delta N^J (\rho^I\, g)_J 
	= \partial^I \Delta \gamma - \commutator{\Delta \gamma}{\tilde{N}^I}\,.
	\end{equation}
Including result~\eqref{eq:Del_rho_line_2} for the second line, the change of $ \hrho^I $ is therefore 
	\begin{equation}
	\Delta \rho^I = \partial^I \Delta \gamma +  \tilde{N}^J \big((\partial^I \Delta \gamma - \Delta \rho^I)\, g \big)_J\,.
	\end{equation}
Obviously, this equation is solved by 
	\begin{equation} 
	\Delta \rho^I = \partial^I \Delta \gamma\,.
	\end{equation}
This concludes the derivation of the transformation rules for the RG functions under a $ U $-rotation of the bare sources.

\subsection{Sample calculation of the \label{app::NI}$ N^I $ counterterm}
Here we outline the SM computation of the left-handed quark contribution to $ N^I$ to 1-loop order in the gaugeless limit.
We begin by computing the 1-loop vertex corrections between $ Q $ and $ a_\mu\,$ in the limit of vanishing $ a_\mu $ momenta.
These are renormalized by the field-strength renormalization of the external quark legs and the current counterterm, $ N^I $. 

For the up-type quark loop, the amplitude is 
	\begin{equation}
	\begin{split}
	\mathcal{A}^P_{1,\mu} &= \mu^{2\epsilon} \!\! \int \dfrac{\dd^d k}{(2\pi)^d} (-i y_u) \dfrac{-i \slashed k}{k^2} (\gamma_\mu t_u^{P}) \dfrac{-i \slashed k}{k^2} (-i y_u^\dagger) \dfrac{i}{(k+p)^2} \\
	&= \dfrac{1}{16\pi^2} y_u t_u^P y_u^\dagger \! \left( -\dfrac{4\pi\mu^2}{p^2} \right)^{\!\! \epsilon} \dfrac{d-2}{d} \dfrac{\Gamma(\epsilon) \Gamma^2(1-\epsilon)}{\Gamma(2-2\epsilon)}\\
	&= \dfrac{1}{32\pi^2 \epsilon } y_u t_u^P y_u^\dagger\, \gamma_\mu\; +\; \text{finite}.
	\end{split}
	\end{equation} 
From the similar loop with an internal down-type quark, the contribution is  
	\begin{equation}
	\mathcal{A}^{P}_{2,\mu} = \dfrac{1}{32\pi^2 \epsilon } y_d t_d^P y_d^\dagger\, \gamma_\mu\; +\; \text{finite}\,.
	\end{equation}
Meanwhile, the counterterms are contained in the tree-level amplitude
	\begin{equation}
	\begin{split}
	\mathcal{A}^{P}_{\ct,\mu} &= K_\epsilon \! \left[\left(\delta^{PQ} +N^{QI} (t^{P}\, g)_I \right) Z_q^{\dagger} \gamma_\mu t^{Q}_q Z_q \right]\\
	&= \dfrac{1}{\epsilon} N^{(1)I}_q (t^P\, g)_I \gamma_\mu + \dfrac{1}{\epsilon} z_q^{(1)\dagger } t^{P}_q \gamma_\mu + \dfrac{1}{\epsilon} t^{P}_q z^{(1)}_1 \gamma_\mu + \ldots\,, 
	\end{split}
	\end{equation}
where the $K_\epsilon$ operator extracts the divergent piece of its argument.
At the 1-loop order the field-strength renormalization goes as 
	\begin{equation}
	z^{(1)}_q\supset -\dfrac{1}{64 \pi^2} (y_u y_u^\dagger + y_d y_d^\dagger ) \,. 
	\end{equation}
For successful normalization, $ K_\epsilon (\mathcal{A}_1 + \mathcal{A}_2 + \mathcal{A}_\ct) = 0\,$, we find
	\begin{equation}
	N_q^{(1)I} (t^P\, g)_I = \dfrac{1}{64\pi^2} \left(y_u y_u^\dagger t_q^{P} - 2y_u t_u^P y_u^\dagger + t_q^{P} y_u y_u^\dagger \right) \; + \; (u\to d) \,.
	\end{equation}
First, we notice that this result is anti-Hermitian, which is a good sign. Next, we need to check that it is possible to find an $ N^{(1)I}_q $, which satisfies the above expression.
To this end, we observe that we can let
	\begin{equation}
	N_q^{(1)I} \hat{g}_I = \dfrac{1}{64 \pi^2} \left(\hat{y}_u y_u^\dagger - y_u \hat{y}_u^{ \dagger}  \right) \; + \; (u\to d)\,.
	\end{equation}
Upon substituting the action of $ t^{P} $ on the Yukawas (Eqs.~\eqref{eq::deltaomegayu} and~\eqref{eq::deltaomegayd}), we see that this is the counterterm for the left-handed quark field at 1-loop.
Similar computations can be done for the other fermions.

\sectionlike{References}
\vspace{-10pt}
\bibliography{References} 	
\end{document}